\documentclass[a4paper, 10pt]{article}

\usepackage{times,xcolor,amsmath,amssymb,epsfig,graphics,graphicx}
\newcommand{\csch}{\textrm{ csch }}
\newcommand{\sech}{\textrm{ sech }}

\newcommand{\be}{\begin{eqnarray}}
\newcommand{\ee}{\end{eqnarray}}

\begin{document}

%%%%%%%%%%%%%%%%%%%%%%%%%%%%%%%%%%%%%%%%%%%%%%%%%%%%%%%%%%%%%%%%%%%%%%%%%%%%%%

%%%%%%%%%%%%%%%%%%%%%

\begin{center}
{\Large \bf Modelling duality between  bound and resonant meson spectra by
means of free quantum motions on  the de Sitter space time $dS_4$ }
\end{center}

\vspace{0.02cm}
\begin{center}
{M.\ Kirchbach$^1$, C.\ B.\ Compean$^2$}\\
{$^1$Instituto de F{\'{i}}sica, UASLP,
Av. Manuel Nava 6, Zona Universitaria,\\
San Luis Potos{\'{i}}, S.L.P. 78290, M\'exico\\
$^2$ Instituto Tecnol\'ogico de San Luis Potos\'{\i},\\
 Av. Tecnol\'ogico S/N col. UPA,\\
Soledad de Graciano S\'anchez near San Luis Potos{\'{i}}, S.L.P. 78437, M\'exico\\
}
\end{center}

\abstract{
The real parts of the complex squared energies defined by the resonance poles of  the transfer matrix of the P\"oschl-Teller barrier, 
are shown to equal the squared energies of the levels bound within the trigonometric Scarf well potential.
By transforming these potentials into parts of the Laplacians describing free quantum motions on the mutually orthogonal open time-like hyperbolic--,  and closed space-like spherical geodesics on the conformally invariant de Sitter space time, $dS_4$, the conformal symmetries of these interactions are revealed. On $dS_4$ the potentials under consideration naturally relate to interactions within colorless two-body systems and to cusped Wilson loops.
In effect, with the aid of the $dS_4$ space-time as unifying geometry, a conformal symmetry based bijective correspondence (duality) between bound and resonant meson spectra is established at the quantum mechanics level and related to confinement understood as color charge neutrality.
The correspondence allows to link the interpretation of mesons as resonance poles of a scattering matrix
with their complementary description as states bound by an instantaneous  quark interaction and to introduce a conformal symmetry based classification scheme of mesons. 
As examples representative of such a duality we organize in good agreement with data  71 of the reported light flavor mesons with masses below $\sim$ 2350 MeV into four-conformal families of particles placed on linear $f_0$, $\pi$, $\eta$, and $a_0$ resonance trajectories, plotted on the $\ell/M$ plane. Upon extending the $\sec^2\chi $ by a properly constructed conformal color dipole potential, shaped after a tangent function, we predict  the masses of 12 ``missing''  mesons. We furthermore notice that the $f_0$ and $\pi$ trajectories can be viewed as chiral partners, same as the  $\eta$ and $a_0$ trajectories, an indication that chiral symmetry for mesons is likely to be realized in terms of parity doubled conformal multiplets rather than, as usually assumed,  only in terms of parity doubled single $SO(3)$ states. We attribute the striking measured meson degeneracies to conformal symmetry dynamics within color neutral two-body systems, and conclude on the usefulness of the de Sitter space-time $dS_4$ as a tool for modelling strong interactions, on the one side, 
 and on the relevance of hyperbolic and trigonometric potentials in constituent quark models of hadrons, on the other. }

\vspace{0.5cm}
{Pacs numbers: {12.39.Jh} {(Nonrelativistic quark models)}, {14.40.Be} {(Light mesons)}, {03.65.Fd} {(Algebraic methods)}, {02.30.Ik} {(Integrable systems)}}

\section{Introduction }
The physics of hadrons is nowadays one of the prolific topics in contemporary research, both experimental and theoretical. The hadronic particles, composed by  quarks, the fundamental matter degrees of freedom  in the gauge theory of strong interaction, the Quantum Chromodynamics (QCD), are subject to studies based both on first principles, among them solving the QCD equations on the lattice, and phenomenological approaches, such as the quark model.
A large amount of data has been accumulated and is  awaiting for more detailed and adequate descriptions, among them data on the light flavor mesons with masses ranging  between $\sim 1400$ MeV and $\sim 2 500$ MeV \cite{PART}.
The fact is that while the first nine mesons of the lowest masses  lie below 1020 MeV, the subsequent nine share the five times smaller mass slot between 1170 MeV and 1370 MeV. This  tendency of dense population of the high lying mass regions by mesons of spins varying from zero to six and their striking  degeneracies with respect to various combinations of the quantum numbers, continues up to $\sim 2500$ MeV, and presents a challenge concerning the possible classification schemes. 
As a simple but illuminating example of the difficulty of the problem, take the following seven nearby mass degenerate mesons, $ \pi_1(1400)$,$\eta (1405)$, $f_1(1420)$, $\omega(1420)$, $f_2(1430)$, $a_0(1450)$, $\rho(1450)$, squeezed within the narrow mass region between $1400$ MeV and $1450$ MeV. Their quantum numbers, denoted by, $I^G(\ell^{PC})$, where $I$ is the isospin, $\ell$ is the total angular momentum (the integer spin), 
while  $P$, $G$  and $C$ are in turn the spatial, the $G$- and the charge conjugation parities,
correspond to, $1^-(1^{-+})$, $0^+(0^{-+})$, $0^+(1^{++})$, $0^-(1^{--})$, $0^+(2^{++})$, $1^{-}(0^{++})$, and $1^+(1^{--})$, respectively.\\

\noindent
{\bf Degeneracies and symmetries.}
To the amount degeneracies are necessarily required by symmetries, various algebraic symmetry schemes can be hypothesized
to explain the above minor mass splittings.
In one of the possibilities one can fix  $CP$ parity and  isospin but allow $\ell$  and the $G$ parity to vary.
Then the $\omega (1420)$ and $f_2(1430)$ mesons can be joint into an isoscalar doublet
of spins $1^{--}$, and $2^{++}$  of natural spatial and opposite $C$ parities,
according to   $0^-(1^{--})$--$0^+(2^{++})$. 
Another similar couple can be formed by the isotriplet  $a_0(1450)$ and $\rho (1450)$ mesons,
corresponding to $1^-(0^{++})$ and $1^+(1^{--})$, respectively. {}For the remaining three mesons, $\pi_1(1400)$, $\eta (1405)$, and $f_1(1420)$,  no match can be found within the group. 
Alternatively, one could see the septet as composed by the 
isotriplet $a_0 (1450)$--$\pi_1(1400)$ mesons of natural $0^{++}$--$1^{-+}$ spatial and equal $C$ parities, 
and the isoscalar $\eta (1405)$--$f_1(1420)$ pair of unnatural $0^{-+}$ and $1^{++}$ spatial, and also equal $C$  parities,
while the three unmatched mesons would be $\omega (1420)$, $f_2(1430)$, and $\rho (1450)$, a scheme in which one has joint mesons of equal isospins, $G$, and $C$ parities and allowed $\ell $ and the $P$ parity to vary. 
Finally,  as a  third option, the four aforementioned mesons could be combined  to parity couples according to,
$0^+(0^{-+})$--$1^-(0^{++})$, and $0^+(1^{++})$--$1^-(1^{-+})$ , corresponding to 
($\eta(1405)$-$a_0(1450)$), and ($f_1(1420)$-$\pi_1(1400)$), respectively.  These parity doublets are constituted each by an isoscalar-, and an isotriplet meson, much alike the $0^+(0^{++})$--$1^-(0^{-+})$  chiral ($f_0 (500)$-$\pi (139)$ ) doublet.

This example shows that mass degeneracy alone is necessary but not sufficient for concluding on a particular symmetry, and that purely algebraic assignments are inevitably plagued by ambiguities, all problems which can be avoided to a large extent by introducing the symmetry through the dynamics. In the following we shall develop a strategy towards the
introduction of such a dynamics. Before, we like to briefly attend to the meson classification scheme of the most frequent use in the literature. \\

\noindent
{\bf Meson Regge trajectories revisited.} Usually, mesons are classified according to  Regge trajectories, 
straight lines on the plane of the total angular momentum ($\ell $) (the integer spin of the particle)  versus the squared invariant mass $(M^2)$ of the type\cite{Gribov},
\begin{equation}
\ell (M^2,t)=\alpha(t)M^2 +\alpha (0), \quad \left[ \alpha (t)\right]=\mbox{MeV}^{-2},
\label{Rgg_can_1}
\end{equation}  
where the argument of the slope, $\alpha (t)$,  is the $t$ channel Mandelstam variable.
Relationships of the type in  (\ref{Rgg_can_1}) appear within string approaches to resonances and some relativistic versions of the quark model \cite{Ebert}, and are valid only grosso modo \cite{Afonin}. 
Indeed, the linearity under discussion requires equidistance between the squared masses, a condition which finds itself notably violated
\cite{ATang}{}, \cite{Lodhi} by the data. For example,
the measured $\left( M^2_\ell -M^2_{\ell -1}\right)-\left( M^2_{\ell -1}-M^2_{\ell -2}\right)$ difference for the $f_0(980)$ meson trajectory,
 and for $\ell =  4$, corresponding to  the $f_4(2300)$-$\omega_3(1945)$-$f_2(1640)$ triad,  is $\sim 0.41555\times 10^6$ MeV$^2$, 
while same difference for the $f_0(1370)$ meson trajectory, and for $\ell =3$, corresponding to the 
$\omega_3(2285)$-$f_2(2010)$-$\omega (1420)$ triad, is $\sim (-0.1365)\times 10^6$ MeV$^2$. Thus, the deviation from the prescribed zero is notable. However, on plots where $\left[M^2\right]$ has been given in units of [GeV]$^2$, 
and where in addition the unit lengths on the  $\ell $ and the $M^2 $ axes have been set equal (Chew-Frautschi plots in the terminology of \cite{Collins}), the deviations from the linearity is camouflaged by becoming less perceivable by mere inspection. The reason behind the deviations of the particle's squa\-red masses from the prescribed linear dependence on their spins has to be looked up in first instance 
in the complicated composite nature of the hadrons. Indeed, the masses of composite systems depend on the masses of their constituents, 
valence quarks and effective gluons in our case, on the one side, and  on the shape of the effective strong potential that keeps the whole complex system together, on the other, all quantities difficult to be figured out with a sufficient accuracy. 
In view of the complexity of the problem, it may appear that searching to improve the Regge classification scheme of hadrons may not be a promising task.
Yet, a lot of more realistic predictions for meson masses  are delivered by constituent quark models, in which hadrons are viewed as bound states within 
well potentials.\\

\noindent 
{\bf Bound and resonant states duality.}
In effect, hadrons in general, and mesons in particular, are described in a twofold way. On the one side they are viewed as virtually exchanged resonances in scattering processes, where they are described as Regge trajectories \cite{Gribov}, and on the other they are treated within the constituent quark model as states bound within well potentials \cite{Valcarce}{},\cite{CK_2010}{}. So far in the literature the two ways have been considered as a rule as complementary to each other and pursued separately 
with the hint on their relevance for different regimes of QCD, a reason for which the question on their possible common root has rarely be posed \cite{Lovelace}.
It is the goal of the present work to provide an answer to precisely this question. Our strategy is  
to describe mesons as resonances transmitted through a barrier whose spatial symmetry is consistent with the spatial  symmetry of  the well potential employed in the description of the bound states and motivate in this manner, by the aid of the dynamics, a symmetry and degeneracy based classification scheme of mesons.

\noindent
{}For the sake of paying tribute to the conformal symmetry of QCD in the ultraviolet regime of QCD, where particles are associated with the scattering matrix poles, we approach our goal by  designing a conformal symmetry respecting duality between levels bound within a well potential and resonances transmitted 
through a barrier. Notice that the relevance of conformal symmetry for hadrons has been addressed by various authors already in the early days of Regge's theory, for example in Refs.~\cite{Freedman}{}, \cite{Domokos}{} \cite{Frazer}, where Regge trajectories with $O(4)$ symmetric poles have been considered.Additional  hints on the possible relevance of the conformal symmetry in the infrared regime of QCD come from the conjectured  duality \cite{AdS} between a weekly coupled string theory and a strongly coupled conformal theory (associated with QCD) at the boundary of the $AdS_5$ space. Experimental data on the possible walking
of the strong coupling towards a fixed value at zero momentum transfer  \cite{Andre} seem to provide further support to the $AdS_5/CFT_4$ conjecture. Within this so called  $AdS_5/CFT_4$ framework, conformal Regge trajectories have been built up recently in  \cite{Conf_Rg}.

As a pair of potentials allowing for the construction of the aforementioned duality we encounter  the trigonometric Scarf
well,$V_{\mbox{Sc}}(\chi)$, earlier considered in \cite{JPAMT2011}, and the  P\"oschl-Teller barrier, ${\mathcal V}_{\mbox{PT}}(\rho)$  \cite{PoTel}, 
\begin{eqnarray}
V_{\mbox{Sc}}(\chi)&=&\frac{\hbar^2c^2}{R^2}\ell (\ell +1)\sec^2\chi,\quad \chi\in \left[-\frac{\pi}{2},+\frac{\pi}{2} \right],
\label{TrigScrf}
\\ 
{\mathcal V}_{\mbox{PT}}(\rho)&=&\frac{\hbar^2c^2}{R^2}\left[(K+1)^2 +\frac{1}{4}\right]\sech^2\rho,\nonumber\\
 \rho &\in& (-\infty, +\infty),
\label{PoschTlbrr}
\end{eqnarray}
where $\ell$  and $K$ stand in their turn for the three-,  and four-dimensional angular momentum values, while $R$ is so far a generic length parameter to be specified below.
In order to make the  conformal symmetry of the $V_{\mbox{Sc}}(\chi)$ well  manifest, we take advantage of the freedom of choice for the coordinates in the associated wave equation, 
which we transform in such a way, that in the new coordinates the well becomes part of the kinetic-energy operator describing inertial quantum motion on  the unique closed space-like geodesic 
of  a four dimensional de Sitter space time, $dS_4$. This geodesic is a three dimensional hypersphere, $S^3$, which
could independently be seen on its own rights as the curved position space of a  $(3+1)$ dimensional  Minkowski space time, contained in $dS_4$, and related to the flat  conformal space-time by a conformal compactification \cite{LuscherMack}, a reason for which the spectrum bound within the well falls as a whole into an infinite unitary representation of the conformal group $SO(4,2)$ \cite{Schr41}.

As a signature for  the conformal symmetry of the bound spectrum of the trigonometric Scarf well  it is commonly accepted to consider the $(K+1)^2$-fold degeneracies of the states in the levels \cite{JPAMT2011}{},\cite{KimNoz}{}.  
The de Sitter space time $dS_4$ has recently attracted attention through the hypothesized relevance of de Sitter special relativity in QCD as advocated in \cite{Pereira},
on the one side, and through its close relationship to the $AdS_5$ space time (fundamental to QCD via the gauge-gravity duality) that allows for $dS_4$ slicing \cite{Karch}, on the other side.\\

The proof of the conformal symmetry of the barrier is a bit more involved. First one  observes that it emerges from the hyperbolic well,
$V_{\mbox{PT}}(\rho)= - \left[(K+1)^2 -1/4 \right] \sech^2\rho$, through the complexification (analytical continuation), of the potential magnitude according to, $(K+1)\longrightarrow i(K+1)$.
Next, the $\left(-\sech^2\rho\right) $ by itself is correlated  with the trigonometric $\sec^2 \chi$ well by a complexification of $\chi$ to $\chi \longrightarrow i\chi$, 
implying that the $\rho$ argument has to change along a hyperbolic geodesic orthogonal to $S^3$. 
With this in mind we transform the wave equation with the hyperbolic well in such a way, that  it gets absorbed by the Laplace operator describing
free quantum motion on $dS_4$ along open time-like hyperbolic geodesics,  orthogonal to $S^3$. To the amount the symmetry of the Laplacian (and  therefore of the ``swallowed'' potential), is same as the isometry of the surface on which it describes the quantum motion, so is the symmetry of the potential in question. 
The isometry group of the $dS_4$ surface is $SO(4,1)$, the maximal non-compact subgroup of the conformal group $SO(4,2)$, thus revealing the conformal symmetry of the hyperbolic P\"oschl-Teller potential. 
In due course, the $\chi$ and $\rho$ variables acquire in their turn  meanings of second polar-,  and first hyperbolic angle parametrizing the $dS_4$ space time, which is a four dimensional hyperboloid of one sheet.
In effect, we find that  the real parts of the complex squared energies corresponding to the  resonance poles of the transfer matrix through the P\"oschl-Teller barrier equal the squared energies of the levels bound within the trigonometric Scarf well and  share  same conformal degeneracies.
In this way, the claimed correspondence is established, which links  the interpretation of
particles as resonance poles of a scattering matrix to their complementary description as states bound within an instantaneous potential.
On the basis of this link we elaborate a classification scheme for mesons according to conformal resonance trajectories on the plane of
the four-dimensional angular momentum, $(K)$, versus the invariant  mass $(M)$, or, equivalently,  on the plane of the  angular momentum, $\ell$, conditioned by, $\ell=0,1,..., K$, with $K=0,1,2,...$, versus the mass $M$. To be specific,  within our suggested scheme we  find as the counterpart to eq.~(\ref{Rgg_can_1}) the following  linear dependence of the total angular momentum on the invariant mass,
\begin{equation}
\ell (M,R)=\alpha (R)M -n-1, \quad \alpha (R)=\frac{R}{\hbar c}, \quad \left[ \alpha (R)\right]=\mbox{MeV}^{-1},
\label{we_ours}
\end{equation}
where $R$ is a length parameter, taken as the $S^3$ radius,  and $n$ stands for the number of nodes in the wave function.
The linearity in the latter equation requires equidistance between the plane masses, which we find fairly well confirmed by data.
Indeed, the $\left( M_\ell -M_{\ell -1}\right)-\left( M_{\ell -1}-M_{\ell -2}\right)$ difference for the $f_0(980)$ meson trajectory, and for $\ell =  4$ is 
$\sim 50$ MeV, while same difference for the $f_0(1370)$ trajectory, and for $\ell =3$ is $\sim (-85)$ MeV.
These numbers are by several orders of magnitude closer to zero than  Regge's squared mass differences discussed above, 
where for the  mesons under discussion they were  encountered  as  $0.4155\times 10^6$ MeV$^2$, and $\sim (-0.1365)\times 10^6$ MeV$^2$, respectively.
Therefore, the linear  dependence of the spin on the mass  advocated in (\ref{we_ours}) meets reasonably  well the linearity criterion for resonance trajectories.
We  apply this scheme to the light flavor mesons with masses below $\sim$ 2350 MeV.
 Specifically, we organize, in good agreement with data,  71 of those  mesons into four-conformal families corresponding to $f_0$, $a_0$, $\pi$, and $\eta$ 
resonance trajectories on the  $\ell=(0,1,..., K)/M$, with $K=1,2,..,5$ planes,  predicting  12 missing states.
Comparison with similar analyzes by other Authors are given in due places in the main body of the text. As long as  the $dS_4$ space-time  has been employed in establishing at the quantum mechanical level the  conformal symmetry respecting correspondence (duality) between bound and resonant meson spectra, we
conclude on its relevance for the modelling of strong interactions, in line  with  ref.~\cite{Pereira}.\\

\noindent
{\bf  Conformal symmetry and color confinement. }
In our study of the properties of the potentials we observe that the $dS_4$ geometric set up  facilitates recognizing a non-trivial link between  conformal symmetry on closed spherical spaces and confinement understood as color charge neutrality. The link is established in formulating effective chromo-statics on $S^3$ and realizing that the $\sec^2\chi $ function shapes there the ${\mathbf E}$-field of a 
color dipole (two-sources) potential defined by means of the Green function of the conformal Laplacian on this space. This finding is suggestive of introducing the notion of a ``geometric confinement''  as a conformal symmetry triggered color neutrality on closed spherical spaces, an  option for the environmental expression of the color neutrality following  from  the color gauge $SU(3)_c/Z_3$ dynamics in QCD. Testing predictive power of the color dipole confining potential is among the goals of the present work.
In due course,  all the potentials under consideration are shown to take their origin from cusped Wilson loops.

\noindent
{\bf Outline of the article.}
The article is organized as follows. The next section is entirely devoted to the quantum mechanical wave equation with the hyperbolic P\"oschl-Teller well potential and its transformation to
free quantum motion along open time like $dS_4$ geodesics for the sake of revealing its conformal symmetry. Section 3 is dedicated to (i) the complexification of  the potential parameter with the aim to transform the well into a barrier, (ii) the consideration of the emerging  complex resonant spectrum.
In section 4 the data on the $f_0$, $\pi$, $a_0$, and $\eta$ meson resonances with masses reaching up to  $\sim $2350 MeV are analyzed and classified according to four conformally symmetric resonance trajectories and in good agreement with the experimental observations.  Also there,
 we upgrade the $\sec^2\chi$  interaction by a potential shaped after a tangent function, motivated in the Appendix B as a conformal symmetry respecting 
color dipole potential, and fit the  parameters of the net potential to the  meson masses lying on the 
$f_0$, $a_0$, $\pi$, and $\eta $ trajectories. In so doing, the masses of 12 ``missing'' resonances are predicted and a realistic description of the mass gaps between the lowest and the next higher lying mesons is achieved. The article closes with a summary and conclusions section, and has two appendices dedicated to the properties of the potentials under investigation.
In the Appendix A  detailed attention is paid to the trigonometric Scarf well, namely, there we show  that the $\sec^2\chi$ function shapes  
on $S^3$, the closed space-like geodesic of $dS_4$, the ${\mathbf E}$ field of a conformal color dipole potential,  a tangent function,
obtained from the Green function of the conformal Laplacian on this surface.  We conclude that conformal symmetry on closed spherical spaces, in  favoring  confinement understood as color neutrality of a system, is suggestive of  a geometric (environmental) definition of confinement as conformal symmetry provoked color neutrality on such closed spaces, or vice versa, $SU(3)_c/Z_3$ color neutrality provoked conformal symmetry on closed spaces. This constructive definition, in predicting the form of the dipole potential generated by a color--anti-color pair, has been tested by data. In the Appendix B  we link the trigonometric and hyperbolic potentials used through the text to Wilson loops with cusps, thus motivating them by fundamental field theoretical principles.

\section{The P\"oschl-Teller well and barrier potentials}

We begin by first introducing the  P\"oschl-Teller (PT) well potential \cite{PoTel}, defined as,
\begin{equation}
V_{\mbox{PT}}(\rho)=\frac{\hbar^2c^2}{R^2}\left[A^2 -\frac{\lambda^2-\frac{1}{4}}{\cosh^2\rho }\right], \quad \rho \in (-\infty, +\infty).
\label{PT_PT}
\end{equation}
It is a non-singular exactly solvable one-dimensional hyperbolic potential which has been extensively studied  within the framework of the super symmetric 
quantum mechanics \cite{susy}, \cite{Wipf}. {}For potential parameters of $\lambda >1/2 $ values,  
$V_{\mbox{PT}}(\rho)$  can have a finite number of bound states whose energies, ${\mathcal E}^2$, 
obtained from solving the associated one-dimensional  stationary wave equation, 
\begin{eqnarray}
\frac{\hbar^2c^2}{R^2} \left[ -\frac{d^2}{d\rho ^2} - \frac{\lambda^2 -\frac{1}{4}}{\cosh^2\rho}+A^2 \right] U(\rho)&=& {\mathcal E}^2 U(\rho),
\label{gnrlz1}
\end{eqnarray} 
and are given by,
\begin{eqnarray}
{\mathcal E}^2&=&\frac{\hbar^2c^2}{R^2}\left[ A^2-\left(\lambda +\frac{1}{2}-n_r \right)^2\right],
\label{gnrlz2}
\end{eqnarray}
where $R$ is a matching length parameter, $n_r$ is the number of nodes of the wave function, and
$A^2$ is a sufficiently large  real arbitrary constant  preventing the squared energy of becoming negative. The precise condition for having $n_r$ bound states within this potential is
\begin{equation}
\lambda+\frac{1}{2}>n_r.
\label{BST_CND}
\end{equation}
Quantum mechanics wave equations with certain hyperbolic potentials, among them the P\"oschl-Teller one, and most important,  for some specific choices of the 
potential parameters, formally allow for presentations as eigenvalue problems of Laplacians on hyperbolic surfaces \cite{EKalnins}.

\subsection{The P\"oschl-Teller potential on the four-dimensional de Sitter space time, $dS_4$ }
It is not difficult to prove that upon setting 
\begin{equation}
\lambda= K+1, \quad K=0,1,2,...,
\label{lmbd5}
\end{equation}
changing the one-dimensional wave function $U(\rho)$ in (\ref{gnrlz1}) according to,
\begin{eqnarray}
U(\rho)&= &\cosh^{\frac{3}{2} }\rho \, \phi_{\bar K}(\rho), 
\label{var_chng}
\end{eqnarray}
introducing the five dimensional pseudo-spherical harmonics,\newline $Y_{{\bar K} K\ell m}(\rho, \chi, \theta, \varphi)$,  as 
\begin{eqnarray}
Y_{{\bar K} K\ell m}(\rho, \chi,\theta,\varphi) &=&\phi_{\bar K}(\rho) Y_{K\ell m}(\chi,\theta,\varphi),
\label{Y_eoneplus4}\\
Y_{K\ell m}(\chi,\theta,\varphi)&=&{\mathcal S}_{n\ell}(\chi)Y_\ell^m(\theta,\varphi),\label{4Dharm}\\
{\mathcal S}_{n\ell}(\chi) &=&\cos^\ell \chi {\mathcal G}^{\ell +1}_n(\sin\chi),\nonumber\\
 n&=&K-\ell.
\label{n4}
\end{eqnarray}
with $ Y_{K\ell m}(\chi,\theta,\varphi)$ and  ${\mathcal G}^{\ell +1}_n(\sin\chi)$  
in turn denoting the four-di\-men\-si\-o\-nal spherical harmonics, and the Gegenbauer polynomials,
the equations (\ref{gnrlz1})--(\ref{gnrlz2}) are transformed into,
\begin{eqnarray}
-{\hbar^2 c^2}\Delta_{dS_{4}} (\rho,\chi,\theta,\varphi) Y_{{\bar K} K \ell m}(\rho, \chi, \theta,\varphi) 
&=&\nonumber\\
{\mathcal E}^2_{dS_4}Y_{{\bar K} K\ell m}(\rho, \chi, \theta,\varphi),&&
\label{LPLS_Dplusone}\\
{\mathcal E}^2_{dS_4} = {\mathcal E}^2 -\frac{\hbar^2 c^2}{R^2}\left( A^2-\frac{3^2}{4}\right).&&
\label{enrg_dSit}
\end{eqnarray}
In (\ref{LPLS_Dplusone}), $\Delta_{dS_{4}}(\rho, \chi,\theta,\varphi)$ denotes the Laplace operator on $dS_{4}$,
\begin{eqnarray}
\Delta_{dS_{4}}(\rho,\chi,\theta,\varphi ) &=&\frac{1}{R^2\cosh^3\rho }
\frac{\partial}{\partial \rho}\cosh^3\rho \frac{\partial}{\partial \rho}
+\frac{ {\mathcal K}^2(\chi,\theta,\varphi)}{R^2 \cosh^2\rho},\label{LBLTR}\nonumber\\
\end{eqnarray}
${\mathcal K}^2(\chi,\theta,\varphi)$ stands for the squared four dimensional angular momentum operator,
given by
\begin{eqnarray}
{\mathcal K}^2(\chi,\theta,\varphi)=-R^2\Delta_{S^3}(\chi,\theta,\varphi)&=&
-\frac{1}{\cos^2\chi} \frac{\partial }{\partial \chi}\cos^2\chi \frac{\partial}{\partial \chi }\nonumber\\
&+&\frac{
{\mathbf L}^2(\theta,\varphi)}{\cos^2\chi},
\label{LB_S3}
\end{eqnarray}
and $\Delta_{S^3}(\chi,\theta,\varphi)$ is the Laplace operator on the three dimensional hyperspherical surface, $S^3$,  of (hyper)radius $R$.
Furthermore,  ${\bar K}$ is the five-dimensional (5D) pseudo-angular momentum value, while
 $K$, $\ell $, and $m$ stand in their turns for the $4D$-, $3D$-, and $2D$ spherical angular momentum values.
These quantum numbers refer to the following reduction chain of the $so(4,1)$ algebra,
\begin{eqnarray}
so(4,1)&\supset& so(4)\supset so(3)\supset so(2)\nonumber\\
{\bar K} &\quad &  K\qquad \qquad\,\,  \ell\quad \quad m.
\label{quantumnumbrs}
\end{eqnarray}
The wave functions $U(\rho)$ in (\ref{var_chng})  are very well known for any arbitrary potential parameter and can be found among others 
in \cite{susy}{},\cite{Gadella2}:
\begin{eqnarray}
U (\rho)=\cosh^{-\frac{a}{2}}\rho P_{n_r}^{-a-\frac{1}{2}, -a-\frac{1}{2}}(i\sinh\rho ),&&\label{wafus}\\
a=\frac{1}{2} +\left( {K} +1\right),&&
\label{ad}\\
P_n^{-a-\frac{1}{2}, -a-\frac{1}{2}}(i\sinh \rho)=\frac{\left(-a+\frac{1}{2}\right)_n} {n!}&&\nonumber\\
\times {_2}F_1\left(-n, n-2a;-a+\frac{1}{2};\frac{1-i\sinh\rho }{2} \right).&&
\label{nr}
\end{eqnarray}
Here,   $_2F_1$ is the hyper geometric function, $(...)_n$ is the Pochhammer symbol, and $\rho$ is the arc of an open time like hyperbolic geodesic on $dS_4$.
In the following the notion of geodesics will be used in reference to the arguments of the quantum mechanical wave functions. 
Correspondingly, the energy in (\ref{gnrlz2}) emerges as,
\begin{eqnarray}
{\mathcal E}^2&=&-\frac{\hbar^2c^2}{R^2}\left[\left( K+1\right) +\frac{1}{2}-n_r \right]^2 +\frac{\hbar^2c^2}{R^2}A^2,
\label{energy}
\end{eqnarray}
while for ${\mathcal E}^2_{dS_4}$ in (\ref{enrg_dSit}) one finds,
\begin{eqnarray}
{\mathcal E}^2_{dS_4}&=&  -\frac{\hbar^2c^2}{R^2}\left({K} -n_r\right)\left({ K}-n_r +3 \right).
\label{Enrg_dS4}
\end{eqnarray}

The form of the Laplace  operator in (\ref{LBLTR}) corresponds to 
a  four-dimensional hyperboloid of one shell, ${\mathbf H}_1^{(4)}$, displayed in Fig.~1, equivalent to $dS_{4}$,  
\begin{equation}
dS_{4}:\quad x_1^2 +x_2^2 +x_3^2+x_{4}^2 -x_0^2=R^2,
\label{dS_def}
\end{equation}
and to the following parametrization of the ambient $M^{4,1}$ space time in global coordinates \cite{Coreans}:
\begin{eqnarray}
x^0&=&R \sinh\rho, \quad \rho\in (-\infty,+\infty),\nonumber\\
x^4&=&R \cosh\rho \sin\chi, \quad \chi\in\left[-\frac{\pi}{2},+\frac{\pi}{2} \right],\nonumber\\
x^1&=&R \cosh\rho \cos\chi\sin\theta, \quad  \theta\in\left[-\frac{\pi}{2},+\frac{\pi}{2} \right],  \nonumber\\
x^2&=&R\cosh\rho \cos\chi\cos\theta \sin\varphi, \quad \varphi\in \left[0,2\pi \right],\nonumber\\
x^3&=&R\cosh\rho \cos\chi\cos\theta \cos\varphi.
\label{chart}
\end{eqnarray} 
The five-dimensional  pseudo-angular momentum value, ${\bar K}$, 
fixes  the eigenvalues of 
${\mathcal C}^2$, one of the Casimir invariants of the $so(4,1)$ algebra, related to the Laplace operator as,
 
\begin{eqnarray}
-\frac{\hbar^2c^2}{R^2}{\mathcal C}^2(\rho,\chi,\theta,\varphi)&=&-\hbar^2c^2 \Delta_{dS_{4}}(\rho,\chi,\theta,\varphi),
\label{pseudoAngMom}
\end{eqnarray}
according to 
\begin{eqnarray}
-\frac{\hbar^2c^2}{R^2}{\mathcal C}^2 (\rho,\chi,\theta,\varphi)\phi_{\bar K} (\rho)Y_{K\ell m}(\chi,\theta,\varphi)=-\frac{\hbar^2c^2}{R^2}&&\nonumber\\
\times {\bar K}({\bar K} +3)\phi_{\bar K}(\rho)Y_{K\ell m}(\chi,\theta,\varphi).&&
\label{EVP}
\end{eqnarray}
Comparison of the latter equation to  refs.~(\ref{LPLS_Dplusone})--(\ref{enrg_dSit}), and (\ref{Enrg_dS4}), reveals the following relation between ${\bar K}$ and $K$, 
\begin{equation}
K-n_r\equiv {\bar K}, \quad K=0,1,2,...,\quad n_r=0,1,2,...
\label{PSDAM_DEF}
\end{equation}
Notice that according to (\ref{PSDAM_DEF}), the ${\bar K}$ quantum number  labels infinite series of states of increasing $K$'s and $n_r$'s but of a fixed difference.
In effect, the P\"oschl-Teller well for the $\lambda$ choice in (\ref{lmbd5}) 
has become a part of the Laplace operator describing free quantum motion along open time-like geodesics on $dS_4$.
In conclusion, in being  transformed  into a part of the Laplacian on the $dS_{4}$ space, whose isometry group is $SO(4,1)$,
the sech$^2\rho$ potential with $\lambda$ in (\ref{lmbd5})  can be considered as $SO(4,1)\subset SO(4,2)$ symmetric.

\begin{figure}
\begin{center}
\resizebox{1.05\textwidth}{!}
{\includegraphics{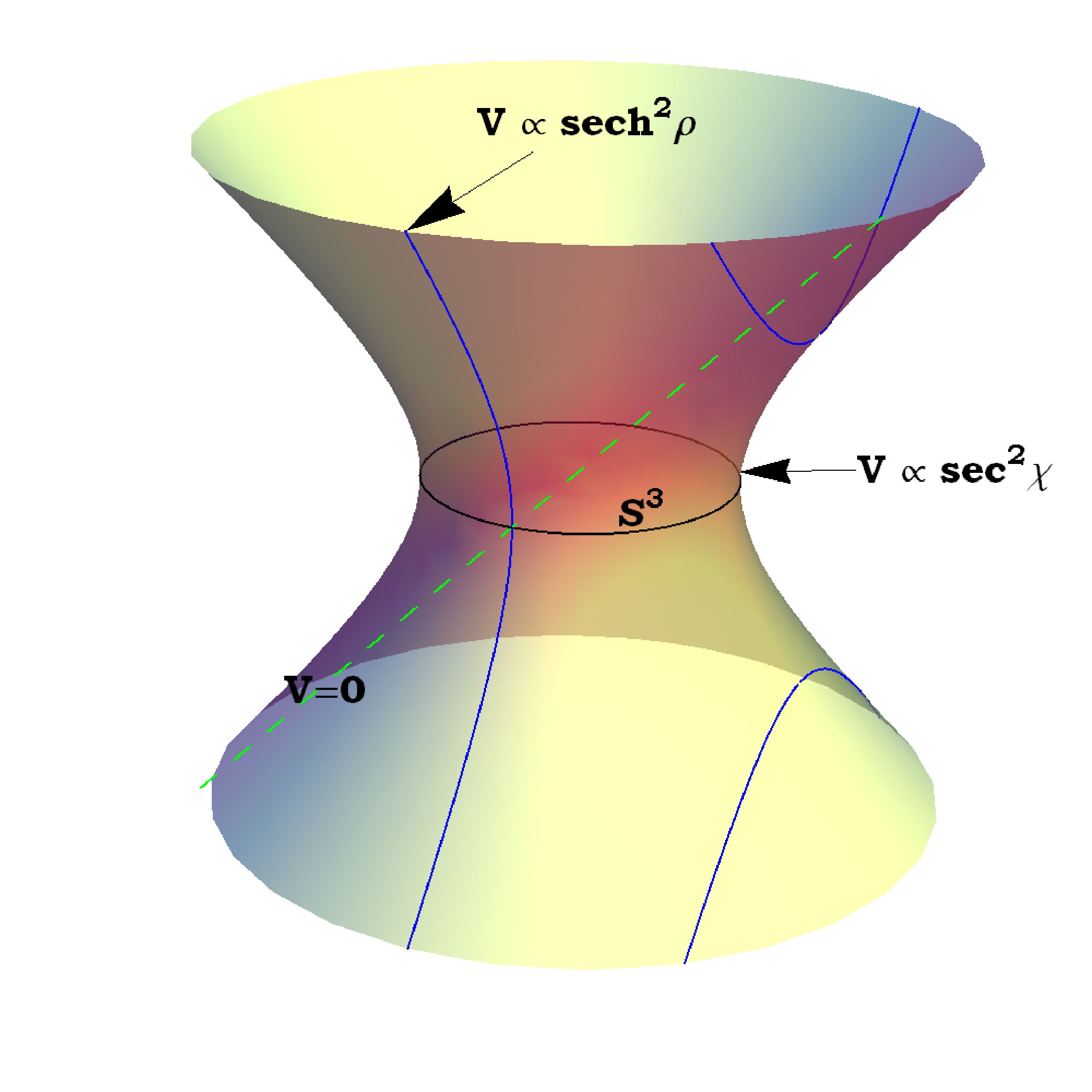}}
\end{center}
\caption{Schematic presentation of the four-dimensional hyperboloid of one shell, the de Sitter space time, $dS_4$, 
 embedded in a five-dimensional Minkowski space-time, $M^{4,1}$,  some of its geodesics, and their associations with one-dimensional potentials.
The equator is a  three-dimensional hypersphere, $S^3$, and a closed space like geodesic. In terms of a Schr\"odinger equation, quantum motion on this space is associated with the one-dimensional 
$V_{\mbox{Sc}}(\chi) $ potential in (\ref{TrigScrf}).
The open hyperbolic  geodesics are three-dimensional  $dS_3$ hyperboloids.  The corresponding one-dimensional potential, 
is $V_{\mbox{PT}}(\rho)$ in (\ref{PT_PT}) for $\lambda =(K+1)$. The straight lines correspond to inertial motions on  three
planes.  The symmetry of the sec$^2\chi $ potential is given by the isometry group $SO(4)$ of $S^3$, while that of 
$\sech^2\rho$ is the $dS_4$ isometry $SO(4,1)$, the respective maximal compact and maximal non-compact subgroups of the conformal group $SO(4,2)$.
The conics obtained through the slicing by planes parallel to the vertical time axis have the structure of three dimensional Lobachevsky space-times, ${\mathbf H}_\pm^3$, 
and describe causal patches on $dS_4$ \cite{Pressley}. \label{Hyperborea}
 }
\end{figure}

\section{ Complexification of the P\"oschl-Teller well to a  barrier,  scattering matrix   and  poles}

In order to transform the well potential in (\ref{PT_PT}) into a barrier it is sufficient to change its sign \cite{Gadella2}{}, which is achieved 
by the following complexification (analytical continuation) of the $\lambda$ parameter in (\ref{lmbd5}):
\begin{eqnarray}
\lambda&\longrightarrow &i\lambda= i\left( K +1\right).
\label{lambda_cmplx_anyD}
\end{eqnarray}
In so doing, one arrives at the following  barrier,
\begin{eqnarray}
 V^{}_{\mbox{PT}}(\rho)&\stackrel{(K+1)\to i(K+1)}{\longrightarrow} &\frac{\left( K+ 1\right)^2 +\frac{1}{4}}{\cosh^2\rho }  +A^2={\mathcal V}_{\mbox{PT}}(\rho).\nonumber\\
\label{PT_PB}
\end{eqnarray}
At the level of the $a$ constant in (\ref{ad}), defining the solutions in (\ref{nr}), the complexification prescription translates  as,
\begin{equation}
a =\frac{1}{2} + K  +1\longrightarrow \frac{1}{2} +i\left(K + 1\right).
\label{key_eqs}
\end{equation}
This is a key relation in the following  as it  turns out to be of crucial importance of finding a geometry on which
the complex poles of the transfer scattering matrix of the P\"oschl-Teller barrier feature the property of carrying same degeneracies as the states in a level bound in the $\sec^2\chi $
 potential in the equation (\ref{ScarfI}) below.    
The effect of the complexification in (\ref{lambda_cmplx_anyD}) on the energy ${\mathcal E}^2$ in (\ref{energy})  is that it changes  from real to a 
complex one, from now onwards denoted by $\left({\mathcal E}^{res}\right)^2$. 
Below we bring the real and imaginary parts of $\left({\mathcal E}^{res}\right)^2$ separately,
\begin{eqnarray}
{\mathcal R}e\,\left( {\mathcal E}^{(res)}\right)^2=\frac{\hbar^2c^2}{R^2}\left[\left(K+ 1\right)^2 -\left( n_r-\frac{1}{2}\right)^2\right] +
\frac{\hbar^2c^2}{R^2}A^2,&&\nonumber\\
\label{Pole_Enrg}\\
{\mathcal I}m\,\left( {\mathcal E}^{(res)}\right)^2= 2\frac{\hbar^2c^2}{R^2} \left(K+1\right)\left(n_r-\frac{1}{2}\right).&&
\nonumber\\
\label{Pole_Width}
\end{eqnarray}

{}From now onwards the $A^2$ constant can be set to zero, $A^2=0$.
Moreover, for the needs of what follows it is convenient to introduce the following notation in terms of a wave vector, $k$, 
defining the complex energy as,
\begin{eqnarray}
\left( {\mathcal E}{}^{(res)}\right)^2&=&\frac{\hbar^2c^2}{R^2}k^2,\nonumber\\
k&=&-i \left[i\left( K+1\right) +\frac{1}{2}-n_r \right]\nonumber\\
 &=&\left[ \left( K+1\right) +i\left( n_r-\frac{1}{2}\right) \right].
\label{k_def}
\end{eqnarray}
In order for the energies in (\ref{k_def}) to be physically observable, one has to show that  $k$ defines a pole of the analytically continued  transmission scattering matrix.

\subsection{Resonances and complex energies}
In order to calculate the scattering matrix of the P\"oschl-Teller barrier, one needs to know the asymptotic behavior of the wave functions
$U(\rho)$ in (\ref{wafus}) for $\rho \to \pm \infty $,  as it follows from  the asymptotic properties of the hyper geometric function
in (\ref{nr}). This behavior is known \cite{AbrStg}{}, \cite{Gadella2} to be given by the following  linear superposition,
\begin{equation}
A_\pm (k) e^{ikx} +B_\pm(k) e^{-ikx}, \quad x=i\sinh\rho,
\label{asmpt_states}
\end{equation}
with $k$ in (\ref{k_def}).
The  scattering matrix connects the asymptotic incoming  with the asymptotic outgoing wave functions, and can be expressed in terms of the  
transfer matrix, ${\mathcal T}(k)$, defined as,
\begin{equation}
\left(
\begin{array}{c}
A_+(k)\\
B_+(k)
\end{array}
\right)={\mathcal T}(k)\left(
\begin{array}{c}
A_-(k)\\
B_-(k)
\end{array}
\right), \quad {\mathcal T}(k)=\left(
\begin{array}{cc}
{\mathcal T}_{11}(k)&{\mathcal T}_{12}(k)\\
{\mathcal T}_{21}(k)&{\mathcal T}_{22}(k)
\end{array}
\right).
\end{equation}
This matrix is expressed in terms of Gamma functions and its explicit  general form  can be found for example in Ref.~\cite{Gadella2}. 
{}For the barrier parametrization of our interest the resulting transmission coefficient, denoted by $T(k)$, can be cast into the following 
form, \cite{Gadella2}
\begin{eqnarray}
T(k)&=&\frac{\mbox{sinh}^2\pi k }{\mbox{cosh}^2\pi k +\mbox{sinh}^2 \pi (K+1)  }.
\label{trnsm} 
\end{eqnarray}
It is easy to cross-check that the $T(k)$ denominator  nullifies for the precise $k$-value given  in (\ref{k_def}), as expected, 
\begin{eqnarray}
\mbox{cosh}^2\pi k +\mbox{sinh}^2 \pi (K+1) =0,&&\nonumber\\
\mbox{for}\quad k=\left[ (K+1) +i\left(n_r-\frac{1}{2}\right)\right],&&
\label{cmplx_zrs}
\end{eqnarray}
thus in confirmation of the anticipated observability of the  complex energies in the equations (\ref{Pole_Enrg})-(\ref{k_def}). Therefore, 
setting   $n_r=$const in (\ref{Pole_Enrg}), one finds out how the real part of the squared complex energy
$ \left( {\mathcal E}{}^{(res)}\right)^2$ looks like at $\rho=0$, i.e. at the  $dS_4$ equator, $S^3$. 
Subsequently, these poles will be at times referred to as P\"oschl-Teller (PT) resonances.

\subsection{Comparison between  the PT resonances and  the states bound  within the trigonometric Scarf well}
The important aspect of the expression for the real part in (\ref{Pole_Enrg}) of the squared resonance energy in (\ref{k_def})  is that it equals, 
modulo an additive constant, the squared energies, here denoted by  $\left({\mathcal E}{}^{(bound)}\right)^2$, 
of the bound states within the  $V_{\mbox{Sc}}(\chi)$  potential in (\ref{TrigScrf}), 
earlier studied in detail in \cite{JPAMT2011} as part of the complete trigonometric Scarf potential.
Without repeating the details of \cite{JPAMT2011},  we notice that  also this potential can be shown to emerge  from the  eigenvalue problem of, $\left[ -\Delta_{S^3}(\chi,\theta,\varphi)\right]$, the Laplace operator  describing free quantum motion on the hyper spherical $S^3$ closed-space like $dS_4$ geodesic, as visible from its definition in (\ref{LB_S3}).
 The corresponding eigenvalue problem reads,
\begin{eqnarray}
-\hbar^2c^2\Delta_{S^3}(\chi,\theta,\varphi) Y_{K\ell m} (\chi,\theta,\varphi) = \frac{\hbar^2c^2}{R^2}&&\nonumber\\
\times  K(K+2)Y_{K\ell m }(\chi, \theta,\varphi) ,&&
\label{RosMor_pre}
\end{eqnarray}
where $Y_{K\ell m }(\chi, \theta,\varphi)$ are the four dimensional hyper spherical harmonics defined in (\ref{Y_eoneplus4})-(\ref{n4}).
Upon re-expressing these  harmonics as, 
\begin{equation}
Y_{Klm}(\chi,\theta,\varphi) \cos\chi = U_{\ell n}(\chi)Y_\ell^m(\theta,\varphi),
\label{var_chng_1}
\end{equation}
and back-substituting in (\ref{RosMor_pre}),
one finds the following one-dimensional stationary wave equation for $U_{\ell n}(\chi)$ solving the $\ell (\ell +1)\sec^2\chi$ potential,
\begin{eqnarray}
\frac{\hbar^2c^2}{R^2}\left( -\frac{{\mathrm d}^2}{{\mathrm d}\chi^2} +\frac{\ell(\ell +1) }{\cos^2\chi} \right)
 U_{\ell n}(\chi)&=&\left({\mathcal E}^{(bound)}\right)^2 U_{\ell n}(\chi),
\nonumber\\
\left({\mathcal E}{}^{(bound)}\right)^2= \frac{\hbar^2c^2}{R^2}(K+1)^2, &\quad&  \chi \in \left[-\frac{\pi}{2},+\frac{\pi}{2} \right].
\label{ScarfI}
\end{eqnarray}
To the amount, the energy at rest equals the invariant mass, the equation (\ref{ScarfI}) can be also read as a mass spectrum,
 \begin{equation}
\left(M^{(bound)}\right)^2\equiv \left( {\mathcal E}^{(bound)}\right)^2= \frac{\hbar^2 c^2}{R^2}(K+1)^2.
\label{mass_bound}
\end{equation}
{}Furthermore, the $n_r$ dependent term in (\ref{Pole_Enrg}) can be absorbed by the real part of the (squared) energy of the resonance, leading to the following
definition of    the real part of the resonance mass square, 
\begin{eqnarray}
\left( M^{(res)}\right)^2 ={\mathcal R}{e}\, \left( 
{\mathcal E}{}^{(res)}\right)^2+
\frac{\hbar^2c^2}{R^2} \left(n_r-\frac{1}{2}\right)^2&&\nonumber\\
=\frac{\hbar^2 c^2}{R^2} (K+1)^2 . &&
\label{Mas_idtfct_res}
\end{eqnarray}
In so doing, the consistency of  these masses  with the masses of the states of equal quantum numbers  when complementary and approximately  treated as bound
in (\ref{mass_bound}) is achieved as,
\begin{equation}
\left( M^{(res)}\right)^2 =\left(M^{(bound)}\right)^2\equiv M^2=\frac{\hbar^2 c^2}{R^2}(K+1)^2,
\label{jkey_eqlty}
\end{equation}
where $M$ is the invariant mass.
Recall,  the branching ratios, 
\begin{eqnarray}
K=n+\ell, \,\,\ell =0,1,2,..., K, \,\, m=-\ell,... 0, ...,+\ell,
\label{branching_rule}
\end{eqnarray}
following from (\ref{quantumnumbrs}), and complementing the equation (\ref{PSDAM_DEF}).
In consequence, the energies in each one of the two  spectra in (\ref{mass_bound}) and (\ref{Mas_idtfct_res}) are $\sum_{\ell=0}^{\ell=K}(2\ell +1)=(K+1)^2$-fold degenerate. Such degeneracy patterns, in combination with be it an infinite number of levels, for the case of the bound spectrum, or the infinite number of  poles, for the case of the resonance trajectories,
are a typical signature of a conformal symmetry \cite{KimNoz}.

\begin{figure}
\begin{center}
\resizebox{0.95\textwidth}{!}
{\includegraphics{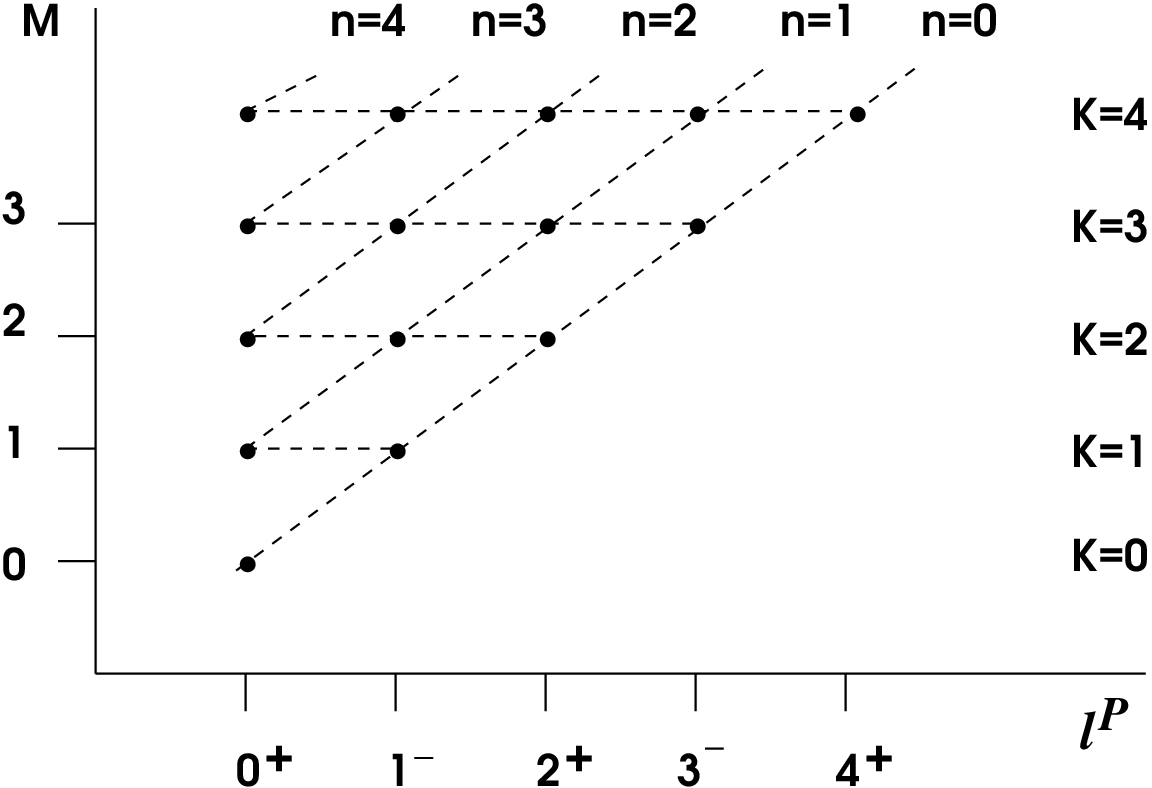}}
\end{center}
\caption{ Schematic presentation on the  $\ell^P/M$ plane (with $P$ standing for spatial parity) of a linear conformal  P\"oschl-Teller (PT) resonance trajectory.
Here, the mass is given in dimensionless units,  
$\left[ \mbox{GeV}\right] \left[ R\right]/\left[ \hbar c\right]$, for simplicity, where 
$\left[ R\right]$=fm. 
The horizontal straight lines mark  finite dimensional representations of the four dimensional rotational group $SO(4)$ corresponding to the levels of the trigonometric Scarf well, their principal quantum number, the four-dimensional angular momentum  $K$,  being given to the very right. 
The ordinary angular momentum content of the $SO(4)$ levels, defined by the branching rule in eq.~(\ref{branching_rule}), 
and for natural parities $(P)$,  can be read off from the horizontal axis. 
The diagonal lines correspond to constant $n$ values and join resonances defined by  the PT barrier according to
(\ref{ellRgga}). The spectrum it is totality represents  a linear  conformal PT resonance trajectory, i.e. the trajectory  with $SO(4)$ symmetric poles in (\ref{KRgg}).
\label{illstr_CS}}
\end{figure}
The issue is that because the isometry of the $S^3$ hypersphere is $SO(4)$, the maximal compact subgroup of the conformal group $SO(4,2)$, the wave functions of the bound states are these very same  ultra spherical harmonics,
$Y_{K\ell m}(\chi,\theta,\varphi)$, previously defined in (\ref{4Dharm}). For this reason, the $\sec^2\chi $ levels are  labeled by the four dimensional angular momentum, $K$ as a principal quantum number, and so are the resonances after the $n_r$=const ``slicing'' in the higher dimension. 
Moreover, because on the other side $S^3$ also represents the curved position space of a Minkowski space time, ${\mathcal M}^{3,1}\in dS_4$, obtained by conformal compactification of the conformal $(4+2)$ dimensional  space time \cite{LuscherMack}, the spectrum as a whole, displayed in in Figure~\ref{illstr_CS},  
falls into an infinite unitary representation of the conformal group, $SO(4,2)$ \cite{Schr41}.
In this way, the spectra in (\ref{jkey_eqlty}) simultaneously implement the conformal symmetry of the 
$\ell (\ell +1)\sec^2\chi $-well potential problem on the one side, and of the real parts of the squared complex energies corresponding to the poles of the transfer matrix of the $[(K+1)^2+1/4]\sech^2\rho$ 
barrier, on the other. This scheme provides the scenario for the conformal symmetry based  classification scheme of mesons, considered in the next section.

\section{P\"oschl-Teller (PT)  resonance  trajectories for high-lying light-flavor mesons and conformal symmetry based systematics }

The bound and resonance states spectra  in the above equations (\ref{mass_bound}), (\ref{Mas_idtfct_res}), and (\ref{jkey_eqlty}) 
are in reality  linear in the  orbital angular momentum, $\ell$, according to,
\begin{equation}
M^{(bound)} =M^{(res)}\equiv M        =\frac{\hbar c}{R}(K+1)=\frac{\hbar c}{R}(n+\ell +1),
\label{linear_spctr}
\end{equation}
where $M$ is the invariant mass.
These equations can be viewed in a twofold way, 
as a linear dependence of the four-dimensional angular momentum on the invariant mass,
\begin{eqnarray}
K (M,R)= \alpha (R) M -1, \quad K=0,1,2,..., && \alpha(R)=\frac{R}{\hbar c},\nonumber\\
\quad \left[ \alpha(R)\right]&=&\mbox{MeV}^{-1},
\label{KRgg}
\end{eqnarray}
or, as a linear dependence of the total angular momentum, $\ell$, on both the  mass and $n$, the number of nodes in the wave function,
\begin{eqnarray}
\ell (M,R)= \alpha (R) M  -n- 1, && \ell=0,1,2,..., \quad \ell +n=K,\nonumber\\
 K&=&0,1,2,...
\label{ellRgga}
\end{eqnarray} 
{}For $n$ taking all the allowed values from $0$ to $K$, the  $SO(4)$ levels of the Scarf well are recovered, while each $n$=const value defines a linear  resonance trajectory corresponding to the P\"oschl-Teller barrier.  In unfolding the dependence in (\ref{KRgg}) on the $\ell /M$  plane according to (\ref{ellRgga}), and assuming  $CP$ conservation, resonance trajectories of the type 
given in Figure \ref{illstr_CS} are obtained. They  fall into infinite dimensional unitary representations of the conformal group, a reason for which such trajectories will also  be termed to as conformal  P\"oschl-Teller (PT)  resonance trajectories.\\

\noindent
As already discussed in the introduction, dependencies of the angular momentum on the invariant mass are generally known as Regge trajectories although
the canonical Regge trajectories refer to a linear dependence of the angular momentum on the squared invariant mass of the type \cite{Gribov},
\begin{equation}
\ell (M^2,t)=\alpha(t)M^2 +\alpha (0), \quad \left[ \alpha (t)\right]=\mbox{MeV}^{-2},
\label{Rgg_can}
\end{equation}  
where the argument of the slope, $\alpha (t)$,  is the $t$ channel Mandelstam variable.
The dependence in (\ref{Rgg_can}) appears within string approaches to resonances. 
We here instead have taken the path of quantum mechanics with the aim to design a correspondence  between bound and transmission resonant states. As it will become clear in due course, data on meson excitations support pretty well the resonance trajectories in
(\ref{KRgg}){}-(\ref{ellRgga})
which we occasionally will also  term to as ``quantum mechanical'' resonance trajectories.
Moreover, our prediction in (\ref{KRgg}) on the $O(4)\subset SO(4,2)$ symmetry of the aforementioned trajectories  is consistent with same symmetry
shared by Regge trajectories of  $O(4)$ symmetric poles, earlier considered by various authors especially in scattering of  particles of equal masses and in the $u$ channel, where the four-dimensional angular momentum is conserved  \cite{Freedman}{}, \cite{Domokos}{}, \cite{Frazer}{}.  
However, while in the latter works the symmetry has been of purely algebraic nature, we here back it up by the conformal potential dynamics.   
It is to be noticed that the  Laplacian in (\ref{LB_S3}) leads to spatial meson wave functions 
describing the excitation modes of a rigid rotor, i.e. of the rotational  modes of two bodies at fixed time-independent  (rigid) distances from their center of mass. At the present stage of development of our model (to be refined below) the nature of the two bodies can not be uniquely specified and can only be conjectured. Some indirect hints are provided  through  the shape  of the potential in the equivalent one-dimensional 
wave equation in (\ref{ScarfI}). At small angles, the $\sec^2\chi $ series expansion gives the Harmonic Oscillator, 
\begin{eqnarray}
\frac{\hbar^2c^2}{R^2}\ell (\ell +1)\sec^2\chi &\approx& \frac{\hbar^2c^2}{R^2}\ell (\ell +1)\left[1 + \chi^2 + {\mathcal O}\left 
(\chi^{2n}\right) \right],\nonumber\\
 n &\geq& 2,
\label{sec2_expansion}
\end{eqnarray}
a circumstance that makes it  more  suitable  for simulating  interactions between more complex meson constituents than the pure 
quark-anti-quark pairs (typically kept together by a stringy potential) thus in principle admitting for, say, unconventional mesons composed by 
$(q\bar q)$- and/or $(q\bar q)G$ ''molecules'' in the spirit of \cite{Jaffe}.
We shall discuss this point in a due place  below. 
 
\noindent 
Linearities as those in (\ref{KRgg})-(\ref{ellRgga})  require eq\-ui\-dist\-ance between the masses, 
while the linearity in (\ref{Rgg_can}) requires eq\-ui\-dist\-an\-ce between the squared masses.
While the first condition finds itself confirmed by data to a good accuracy, the second one is notably violated
especially on  plots in which $M^2$ has been given in units of MeV$^2$ {}(see  \cite{ATang}{}, \cite{Lodhi}).  
In the next section we show that the conformal quantum mechanical PT resonance trajectories  turn out to be pretty  well suited for the description of the high-lying light flavor mesons with masses above $\sim $1400 MeV and below $\sim$2350 MeV.

\subsection{Classifying reported  light-flavor  mesons according to linear PT trajectories}

\begin{figure}
\begin{center}
\resizebox{1.01\textwidth}{!}
{\includegraphics{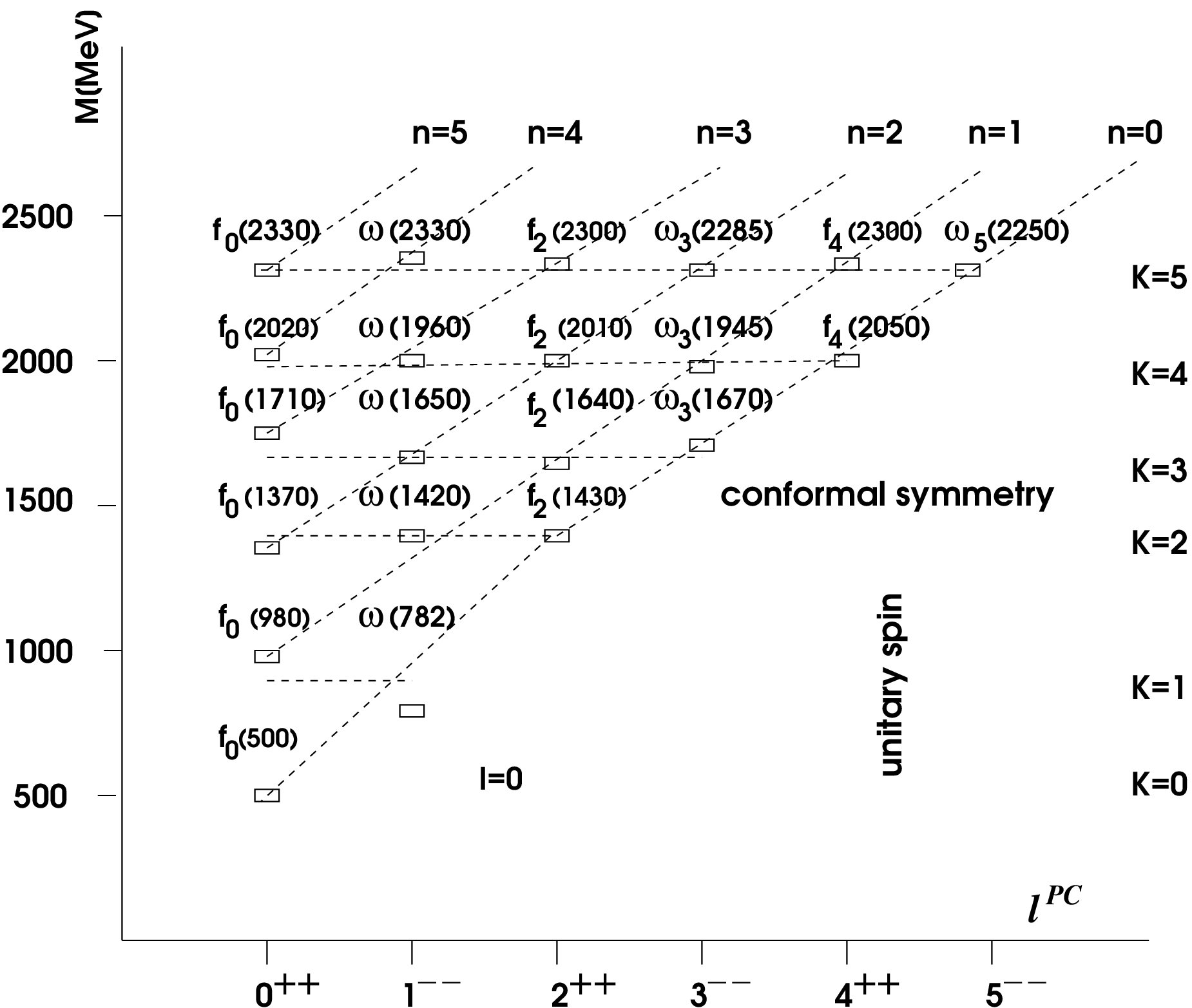}}
\end{center}
\caption{Classification of the isoscalar natural parity $f_0$-meson type  excitations according to a conformal resonance trajectory of the kind displayed  in 
Fig.~\ref{illstr_CS}. Isospin and  $CP$ parity are same for all mesons on the trajectory. The masses  
are given in units of MeV for the sake of having better match with the notations of the mesons, where the number in the parenthesis  is representative for the world average of the measured mass in MeV. 
 The data are taken from \cite{PART}. All $\omega $ mesons from the $K=4$ and $K=5$ poles have been taken from the list of ``Other light mesons''.
 Other notations as in Fig.~\ref{illstr_CS}.
The leading trajectory reveals the most significant deviation from the straight line but this falls into the low-energy and low-$K$ values region,  
where unitary and conformal symmetries interfere.
However, above $\sim$1300 MeV, and for  $K\geq 2$ the trajectories notably straighten.
This is the scale from which onward in our opinion  the conformal symmetry starts  holding valid to a good accuracy.
The $\omega$ meson trajectories, usually displayed separately \cite{Ebert}, are within this scheme  a subset of the $f_0$ trajectories.
The $f_0$ and $\pi$ trajectories  can be viewed as chiral partners (see discussion in Fig.~\ref{Reggepizero}).
\label{eqd1}}
\end{figure}

\begin{figure}
\begin{center}
\resizebox{1.01\textwidth}{!}
{\includegraphics{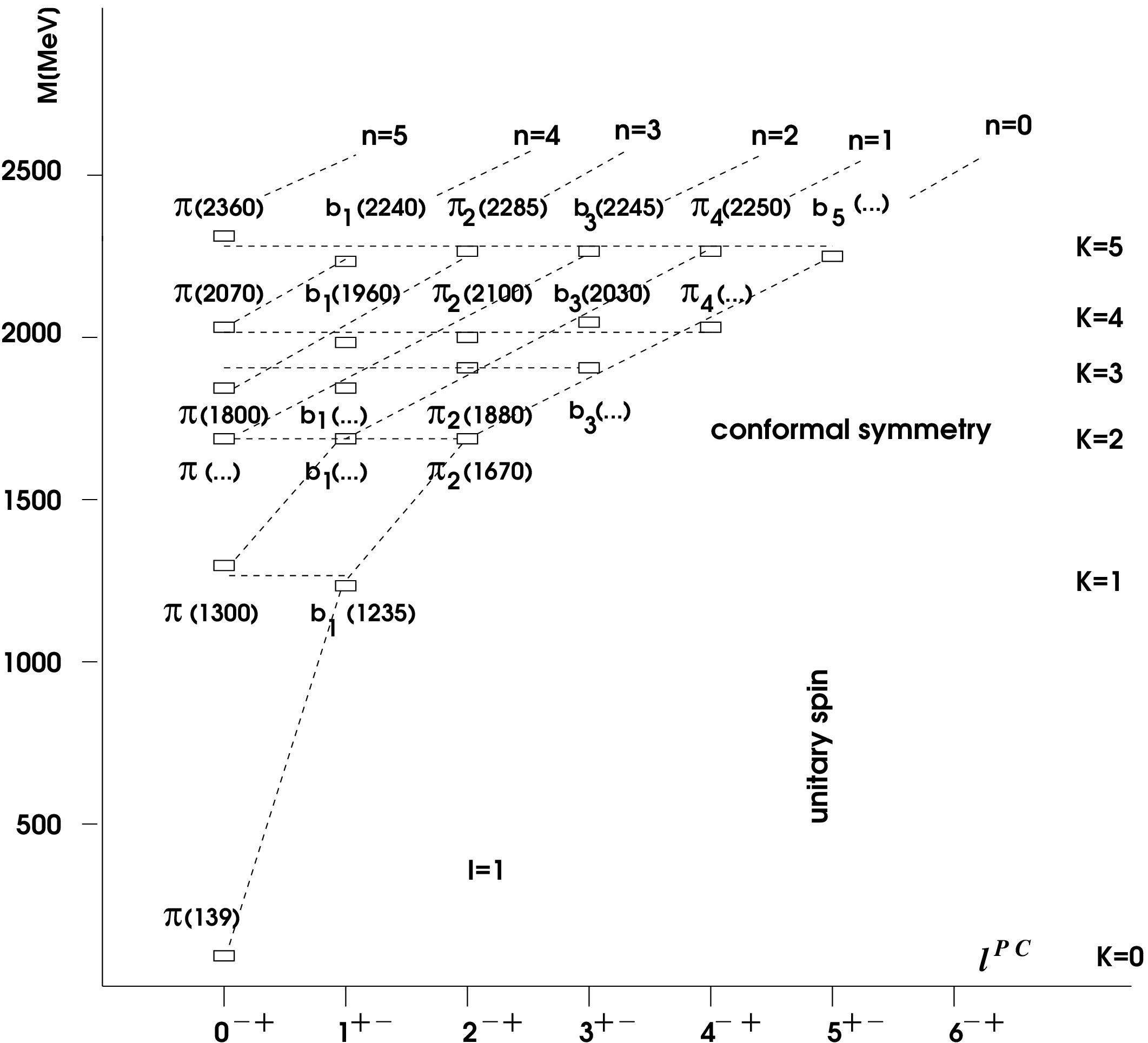}}
\end{center}
\caption{Classification of the isotriplet unnatural parity $\pi $-meson type excitations according to a conformal resonance trajectory of the kind displayed  in
Fig.~\ref{illstr_CS}. Data taken from \cite{PART}. Other notations same as in Figure \ref{eqd1}. Two pions and three $b$ meson states,
denoted by $\pi (...)$ and $b_\ell (...)$  with $\ell=1,3,5$, respectively,  are missing for the completeness of the $SO(4)$ poles  on the figure.
Above $\sim$1800 MeV the trajectories start notably straightening. To the amount within the $\sigma$ model  the $f_0(500)$ could be considered as the chiral partner to $\pi (139)$, the $f_0$ and $\pi$ trajectories could be viewed as chiral partners too.  Notice that above $\sim$ 2000 MeV, the masses of the natural and unnatural parity $SO(4)$  poles with $K=4$ and $K=5$  from the respective $f_0$ and $\pi$ trajectories, become very close.
Admittedly, with the exception of $\pi_2(2100)$, all the other resonances from the $K=4$ and $K=5$ poles have been taken from the list of ``Other light mesons''.
At lower masses, the pole  splittings remain still significant. 
The scale of chiral symmetry restoration for the light flavor mesons in the stronger sense (see discussion in the text) is expected  to happen above 
$\sim$2000 MeV.   
\label{Reggepizero}}
\end{figure}

\begin{figure}
\begin{center}
\resizebox{1.01\textwidth}{!}
{\includegraphics{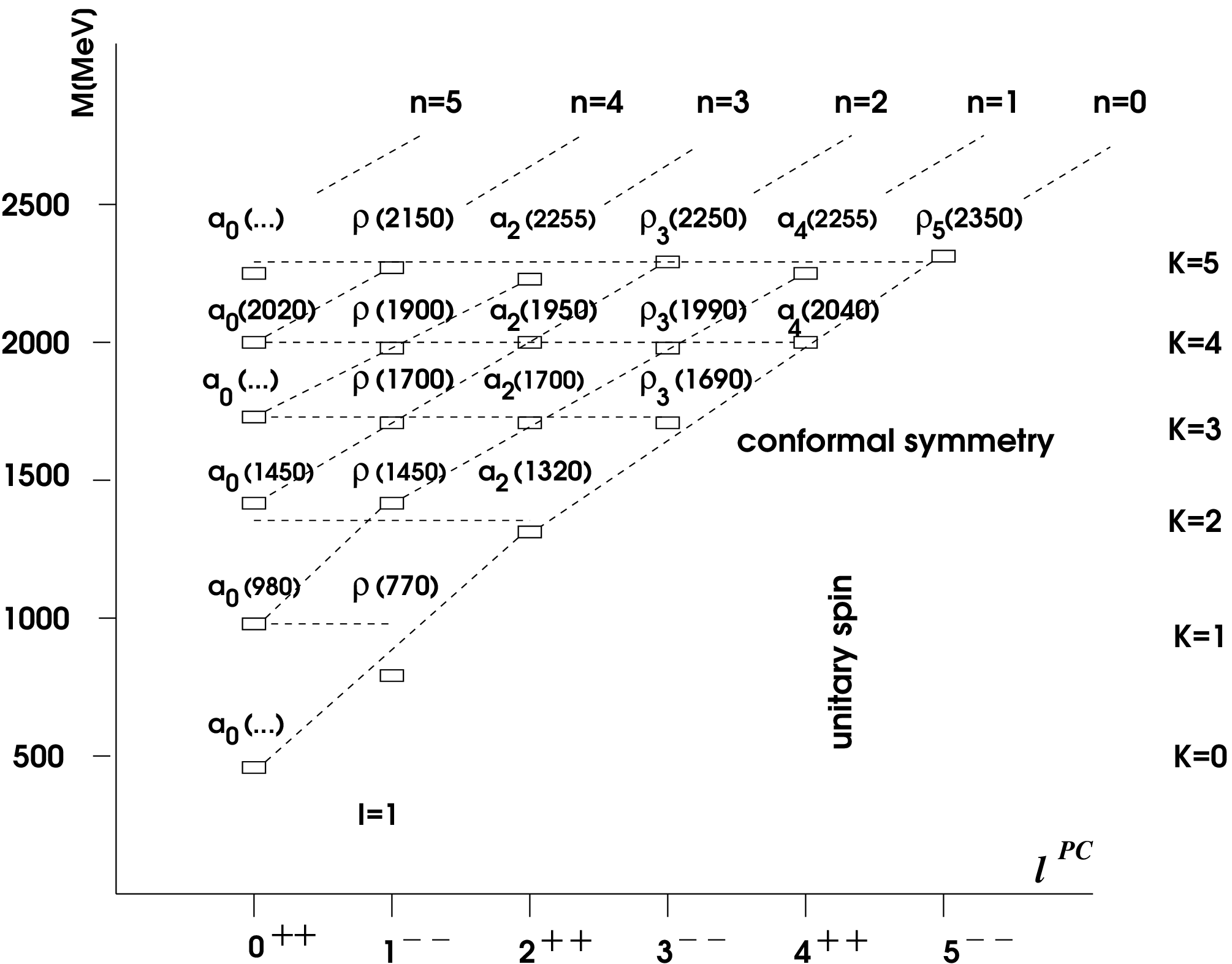}}
\end{center}
\caption{Classification of the isotriplet natural parity $a_0$ meson type excitations  according to a conformal  resonance trajectory of the kind displayed  in
Fig.~\ref{illstr_CS}. Data taken from \cite{PART}.
There are three $a_0$ states,  denoted by $a_0(...)$,  missing for the completeness of the spectrum. Other notations as in Figure \ref{eqd1}.
The higher lying trajectories are straight lines to a reasonable accuracy. In parallel to the $f_0$ case,  we here consider the trajectories of the $\rho $ meson (the triplet partner to the isoscalar $\omega$ meson) 
as a subset of the $a_0$ trajectories and not as usually displayed \cite{Ebert}, as independent trajectories.
The figure shows that above $\sim$1400 MeV the  trajectories on the $\ell/M$ plane are to a good accuracy straight parallel lines.
Similarly to the $f_0$ and $\pi$ trajectories, also the $a_0$ and $\eta$ trajectories are likely to be  chiral partners. 
Exactly in same regions as for the $f_0$ and $\pi$ trajectories, namely around $\sim$2000 MeV and $\sim$2300 MeV,
the masses of the natural and unnatural parity $SO(4)$  poles with $K=4$ and $K=5$  get very close.
However, all the $a_\ell $ mesons in this region have been taken from the list of ``Other light mesons''.
At lower masses, the pole  splittings remain still significant. 
Also for the chiral pair of the $a_0$ and $\eta$ resonance trajectories, the chiral symmetry restoration in the stronger sense (discussed in the text) 
is likely to  happen above $\sim$2000 MeV.  
\label{eqd2}}
\end{figure}

\begin{figure}
\begin{center}
\resizebox{1.01\textwidth}{!}
{\includegraphics{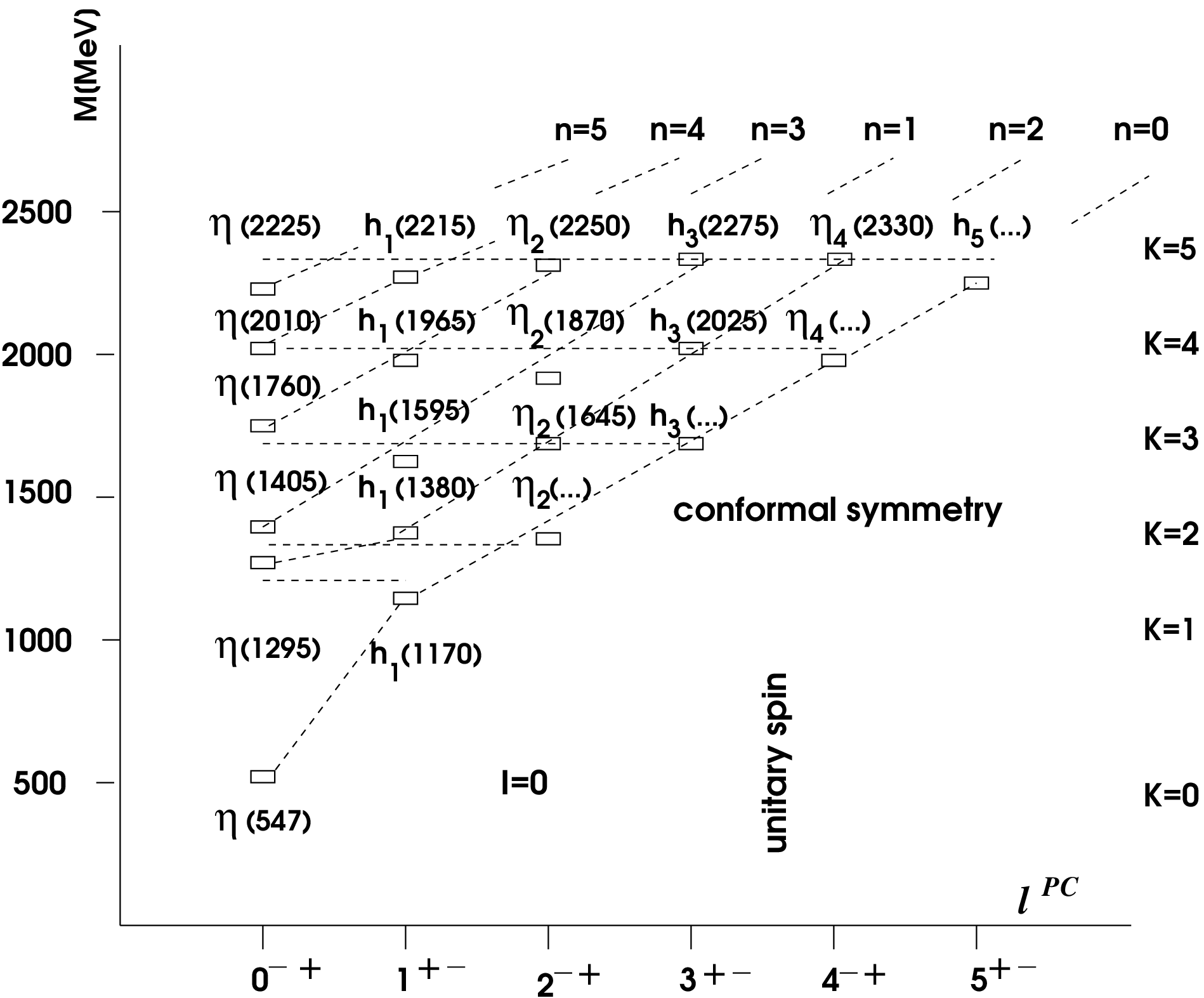}}
\end{center}
\caption{Classification of the isoscalar unnatural parity $\eta $ meson type excitations according to a conformal resonance trajectory of the kind  displayed  in
Fig.~\ref{illstr_CS}. Data taken from \cite{PART}.
The  first $\eta _2$,  $\eta_4$, and $h_3$ and $h_5$ states are ``missing''  for the completeness of the spectrum.
 Other notations as in Figure \ref{eqd1}.
The higher lying  trajectories are straight lines to a reasonable accuracy. 
The figure shows that above $\sim$ 1500 MeV the trajectories on the $\ell/M$ plane are in addition to a good accuracy parallel lines.
In the $K=4$ and $K=5$ levels, with the exception of the $\eta (2225)$ and $\eta_2(1870)$ mesons, all the other resonances have been taken from the list of ``Other light mesons''. This trajectory can be viewed as the chiral partner to the $a_0$ one (see discussion in Figure~ \ref{eqd2}).
  \label{eqd3}}
\end{figure}

In the present section we analyze data \cite{PART} on light-flavor isoscalar--,  and isotriplet mesons of both natural and unnatural parities.
Besides the full meson listings,  use of the  ``Other light mesons'' list has been made.
We classify on the $\ell/M$ plane 71 of those mesons with masses below $\sim$2350 MeV according to four conformal resonance trajectories of the type in
(\ref{ellRgga}),  with ground states corresponding to the $f_0$, $\pi$, $a_0$, and $\eta $ mesons of lowest masses,
displayed on the respective Figures
\ref{eqd1},   \ref{Reggepizero}, \ref{eqd2}, and \ref{eqd3}. Each trajectory has been assumed to be characterized by 
a set of fixed internal quantum numbers corresponding to isospin and $CP$ parity. 
The overall impression one gets by inspection of the four  figures is that with the increase of the mass, the mesons start adjusting better and better 
to straight lines. The mesons with the lowest  masses are commonly organized  into the $SU(3)$ (not shown here)  flavor octet structures although the $K=1$  $SO(4)$ 
doublets, $(\ell =0^\pm, \ell=1^\mp )$,  are already present also in this region for all four meson families. 
Such doublets are constituted by the $[f_0(980)$--$\omega(783)]$, 
$[a_0(980)$--$\rho(770)]$, $[ \pi (1300)$--$b_1(1235)]$,  and $[\eta(1295)$--$h_1(1170)]$ pairs, all of which  carry the correct quantum numbers 
required by the conformal trajectories, although their splittings from the respective ground states, and  the next excited states  do not follow the theoretically  predicted equidistance. In the next subsection we shall attend to this issue in greater detail.
Namely, there we shall show that a data fit with a conformal extension of the $\sec^2\chi $ potential can account for this effect.   
On total, we accommodate  71 observed and predict 12 ``missing'' mesons. However, it needs to be said that the particle's widths do not follow the pattern in (\ref{Pole_Width}) prescribed by the external conformal symmetry and a deeper insight needs to be gained into the internal hadron dynamics to understand their behavior in greater detail.\\

\noindent
One more intriguing observation we have to report concerns the relevance of the (admittedly approximate) chiral symmetry of the QCD Lagrangian
within the light flavor sector of the  $u$ and $d$ quarks from the first generation, i.e. of its symmetry
with respect to continuous transformations of the type 
$ \left[\exp \left(i\frac{\vec \alpha \vec \tau}{2} \otimes {\mathbf 1}_4\right)\exp \left(i\frac{\vec\beta \vec \tau}{2}\otimes  \gamma_5\right)\right]
\left[ \chi \otimes \psi\right]$
where the components of  $\vec\tau$ are the Pauli matrices in isospin space,  $\vec{\alpha}$, and $\vec \beta$  are parameter triplets,
 $\psi $ is a  Dirac field, while $\chi$ stands for the fundamental doublet of the $SU_I(2)$ group of isospin. The symmetry can be described by means of the well known linear $\sigma$ model.  This (Lagrangian) model  is $SO(4)$ symmetric, though in isospin space, and based on a four-dimensional  massless $(1/2,1/2)$ iso-mul\-ti\-plet 
that  contains next to an isoscalar $f_0$ meson, also an isovector $\pi $ meson.  The latter quadruplet  is placed within a double well potential allowing for a non-vanishing (anomalous) vacuum expectation value of the isoscalar field, in consequence of which it becomes possible for the scalar meson to acquire a finite mass,  while the isovector meson remains massless. Within this ``spontaneously broken''  mode of chiral symmetry realization, 
a massive isoscalar $f_0$ and a massless isovector  $\pi$  mesons are considered as ``hidden''  $0^\pm$ chiral partners.   
If the theory were to be free from anomalous expectation values,  chiral symmetry had to be realized instead in the  manifest Wigner-Weyl mode which would require
the $f_0$ and $\pi$ mesons to be of equal masses. 
In general, chiral symmetry requires the duplication in parity of particles of equal spins, and allows their isospins to be distinct by up to one unit. With this in mind, it is now  instructive to compare the    
 $f_0$ and $\pi$ trajectories and to check as to what extent the mesons placed on them can be viewed as chiral partners. 
Such a partnership in the sense of a ``hidden'' chiral symmetry suggests complete duplication of the particles participating the  $SO(4)$ poles of natural parities on the isoscalar $f_0$ trajectory by particles participating the $SO(4)$ poles of unnatural parities on the isovector $\pi$ trajectory, however without demanding coincidence of their masses, something which is fully  satisfied for the lower lying $K=1$ and $K=3$ poles. However, for the highest $K=4$ and $K=5$ poles, where only the $b_5$ partner to the $\omega_5 (2250)$ state is missing, the natural parity $SO(4)$ poles from the $f_0$ trajectory present themselves almost mass-degenerate with the unnatural parity $SO(4)$ poles from the pion trajectory, and more in accord with the  Wigner-Weyl  mode of chiral symmetry realization.
Similarly, also the $a_0$ and $\eta$ trajectories could be viewed as chiral partners,  and the above discussion would equally well apply to them too.
In this way the classification scheme under discussion relates to chiral QCD dynamics.
However, the  chiral symmetry of the canonical QCD Lagrangian in the $(1+3)$ dimensional Minkowski space time does not require and can not explain the observed degeneracy  correlations among  the masses of parity couples characterized by  different  spins. This latter phenomenon requires for its explanation the involvement of a symmetry bigger than the traditional chiral one,  which we here identify as the conformal symmetry. On the other side, chiral symmetry in conformal space-time would refer to parity duplicated irreducible unitary $so(2,4)$ representation spaces, to which the  conformal families of trajectories displayed  in the Figures 3-7 have been set in correspondence.
In this way, the spectra in the Figures 3-7 could be indicative of chiral dynamics in conformal space time. 
Within this context, the masses of the mesons in the two poles of the highest $K$ values observed, located in the region between $\sim$2000 MeV to $\sim$2350 MeV, could indicate the scale at which the restoration of the chiral symmetry from the Goldstone to the Wigner-Weyl mode is likely to take place.

\noindent
The conformal symmetry of mesons has previously been noticed and discussed  also in \cite{Afonin}, \cite{CK_2010}, and \cite{AIP_2012}.
Our analysis, modulo few different assignments, basically because of data upgrades,  is pretty close to that performed by  Afonin in \cite{Afonin}, \cite{Afonin1}.
However, the physical background of our classification, and the conclusions drawn from our analysis, go beyond those in \cite{Afonin}, where the conventional Regge trajectories
in (\ref{Rgg_can}) on the $\ell/M^2$ plane have been
employed. We here have instead first identified a quantum mechanical barrier potential as the culprit behind the conformal patterns of the resonances, 
whose energies are defined by the poles of the associated transmission scattering matrix. Then we could correlate  those patterns  with states bound within a  conformally  
symmetric well  potential. 

\noindent
Finally, a comment is in place on the mesons from \cite{PART} which so far have not been included here.
In first place these are the $f$, $a$, and $\eta$, and $\pi$   mesons of odd spins, and the $\omega$, $\rho$, $h$, and $b$ mesons of even spins.
\textcolor{red}{These groups  of mesons appear  on the $f_1$, $a_1$, $\eta_1$, and $\pi_1$ trajectories.}
These trajectories are of a quality more or less  comparable to the ones analyzed here and have been omitted for the only
sake of not overloading the presentation. 
More seriously, in the energy range under investigation we found the following  11 mesons which dropped out of our suggested systematics:
(i) the one $\rho$ meson, $\rho(1570)$, (ii) the ten $f$ mesons, $f_0(1500)$,  $f_0(2100)$, $f_0(2200)$, $f_2(1270)$, $f_2(1565)$,  $f_2(1810)$, 
$f_2(1910)$, 
$f_2(1950)$, $f_2(2150)$, $f_2(2340)$. In order to gain some  insight into the abundant presence of $f$ mesons in \cite{PART} 
we notice that mesons  like $f_0(500)$, $f_0(980)$, $f_0(1370)$, and $f_0(1710)$ debated in the literature to have same internal structure
(be it quark-antiquark, $(q\bar q)$ \cite{Rischke},  or of a singlet-glueball $(q\bar q)_0G$ \cite{Cheng}) 
appear placed also within our classification scheme on the same trajectory. In view of this observation,
the abundant $f_0$ mesons could be of a structure different from that of the particles  populating the trajectory. Which precisely, remains an open question which can not be answered within the model under discussion. 
Ours is an  algebraic potential model, whose prime purpose is to draw attention and provide a  scenario suitable  for the description of the  striking multiple  degeneracies observed in the meson spectra in terms of a symmetry possibly relevant for the theory, and which does  not allow
one to distinguish between the above two  pictures of microscopic structure.
However, within the context of the discussion after the equation (\ref{sec2_expansion}) above, one should not exclude the option that
the $f_0$ particles on Figure 3 might throughout  be close to a particular type of unconventional mesons. 
One more point that deserves attention concerns the dependence of the suggested classification scheme  on the so far uncertain status of the resonances taken from the list of ``Other light mesons'' in \cite{PART}. First of all we like to emphasize that the scheme employed so far 
does not take into account important aspects of the  internal dynamics. Specifically,
 threshold effects  could frustrate the formation of some of the states predicted by the conformal symmetry,
while many-body effects can cause co-existence of distinct symmetry patterns in the spectra, a phenomenon well documented in nuclear physics
where one can observe coexistence of isolated single-particle and correlated collective rotational excitations  in the spectra of collective nuclei.
Without entering into the technical details of this  phenomenon, which can be extended to apply to the case of our interest too,
we here limit ourselves to notice that the notion of ``partial symmetries'' has been coined in \cite{Leviatan}, where examples of Hamiltonians
adequately describing these properties have been constructed. Our case here is that a clear footprint of the conformal symmetry is seen in the meson spectra. At which scale, through which effects, and to what extent this symmetry could be broken, is a challenging and important issue worth of being investigated in more detailed models of internal meson structure than the one considered here. As to the allegedly  breaking of the conformal symmetry by the dilaton mass, we like to remark that an external scale can but must not necessarily break an algebraic  symmetry, and one has to attend to this question separately by modelling both alternatives and comparing to data. After the equation (\ref{constants_tRM})  we specifically discuss how the  conformal symmetry can be realized in the presence of two external scales, and if the one of them were to be in some way related to the dilaton mass, the latter still could  leave the symmetry intact.

\begin{quote}
In conclusion, the satisfactory  data  classification of the high lying mesons by means of   the conformal symmetry respecting duality between hadronic bound states and
transmitted resonances hints on the relevance of conformal symmetry for strong interaction processes.
\end{quote}

\subsection{Data fit by the (sec$^2\chi $ +tan$\chi$) potential} 
\begin{figure}
\begin{center}
\resizebox{0.91\textwidth}{!}
{\includegraphics{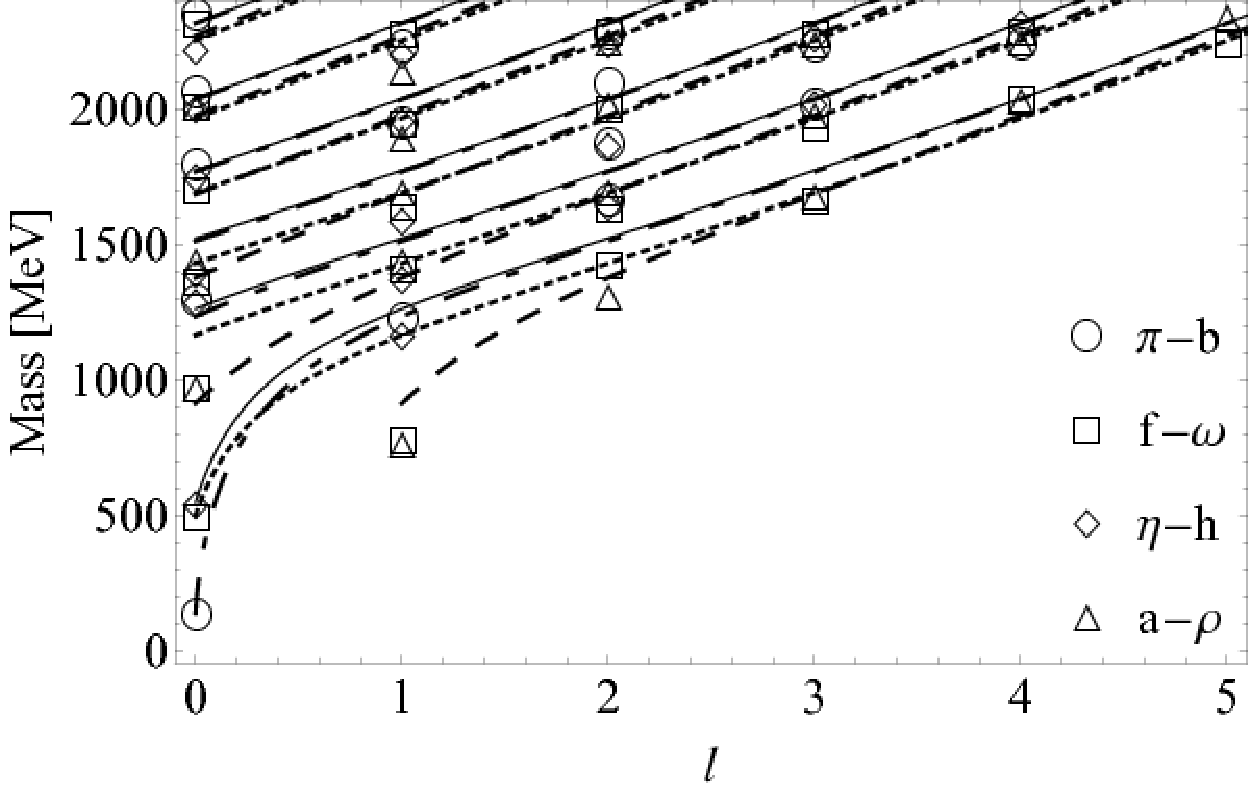}}
\end{center}
\caption{Nine parameter fit by the mass formula in (\ref{energies_tRM}) to the twenty four conformal  trajectories,
corresponding to $\pi(139)$, $f_0(500)$, $\eta (550)$, and $a_0$ resonances.  
The ``missing'' $a_0(...)$ ground state has not been taken into account.
The six  dashed-dotted lines represent our fit to the six pion sub trajectories on the $\ell/M$ plane,  
while the  circles indicate the mass positions of the  observed resonances, as a comparison.
The six solid--, the six  dotted-, and the six  dashed lines, represent in their turn the  fits to the $f_0$, $\eta$ and $a_0$  trajectories, respectively.
The corresponding experimentally observed mesons  located on the latter trajectories are denoted by  squares,  diamonds, and triangles, respectively. The slope parameter $A(R)$ has been kept fixed. The linearity of, and the parallelism among the trajectories above $\sim 1400 $ MeV are well pronounced.
The parity doubling on the  ($f_0$-$\pi$) and ($\eta$-$a_0$) trajectories is also well pronounced. 
\label{allinone}}
\end{figure}

As already observed above, and visualized by the Figures \ref{eqd1}-\ref{eqd3}, the linear $\ell (M,R)$ dependence in (\ref{ellRgga}) between the spins and the masses of the mesons begins applying from about $\sim 1400$ MeV onwards. Below, the deviations of the masses of the $SO(4)$ poles from the equidistance required by (\ref{ellRgga}) are severe and hint on the insufficiency of the  $\sec^2\chi$ potential to correctly capture the dynamics at low energies.
However, a remedy to the problem is provided upon upgrading the $\sec^2\chi$ well by a potential shaped after a tangent function, shown in the Appendix A to represent a conformal symmetry respecting dipole interaction within a color-neutral two-body system.
This  confining color dipole potential is introduced in the equation (\ref{dipole_ctg}), and  its equivalent, calculated from a
 cusped Wilson loop, is given in the equation (\ref{thatis}) in the Appendix B. The potential in (\ref{thatis}) is shaped after a cotangent function and is  related to  (\ref{dipole_ctg}) through a shift of $\chi$  towards $\chi\longrightarrow (\chi +\pi/2)$. 
The net $(\sec^2\chi +\tan\chi)$ potential, denoted by $V^{(b)}_{\mbox{tRM}}(\chi)$, and employed in the data fit in the current section,
is  given in (\ref{RM_V1}), however in dimensionless units.
Then the parameters of the $\left( \epsilon^{(b)}_{\ell n }\right)^2$ spectrum formula in (\ref{RM_V1}),
can be adjusted by least square fit to the masses of the low lying mesons for any one of the  four trajectories. {}For this purpose, we return to the physical units of the energy, $\left[\hbar^2c^2/R^2\right] \left( \epsilon^{(b)}_{\ell n}\right)^2$, and set it equal to the squared  mass, $M^2$,  now expressed in  units of GeV$^2$, arriving  at the following mass formula,
\begin{eqnarray}
\frac{\hbar^2 c^2}{R^2}\left( \epsilon_{\ell n}^{(b)}\right)^2\equiv M^2= A(R)(K+1)^2 -\frac{B(R)}{(K+1)^2} +C,&&\nonumber\\
\label{energies_tRM}\\
A(R)=\frac{\hbar^2c^2}{R^2}, \,\,  \left[ R\right]=\mbox{fm}, \quad B(R)=\left( \frac{b\hbar c}{R}\right)^2, && \nonumber\\
\left[A(R)\right] =\mbox{GeV}^2,\quad \left[ B (R) \right]=\mbox{GeV}^2, \,\,  \left[C\right]=\mbox{GeV}^2,&&\nonumber\\
\label{constants_tRM}
\end{eqnarray}
where $R$ stands for the $S^3$ (hyper)radius.
Contrary to (\ref{ellRgga}), the $\ell (M)$ dependence following  from (\ref{energies_tRM}) is no longer linear.
Notice that the spectrum in (\ref{constants_tRM}) continues respecting the conformal degeneracy patterns of (\ref{ScarfI}) though
it contains next to the first external scale given by the hyper-radius $R$, a second one, the magnitude $B(R)$ of the tangent term.
If the $B(R)$ parameter were to be in some way related to the dilaton mass, then the latter could throughout leave the conformal symmetry intact.
In more technical terms, it can be shown that the Hamiltonian in (\ref{RosMor}) is intertwined  
with the $\chi$ dependent part of the Lapalcian in (\ref{LB_S3}).
The parameter values fitting the data are listed in the Table 1.
The slope parameter, $A(R)$, has been kept fixed for all four trajectories, while the remaining two parameters, $B(R)$ and $C$, were allowed to change from trajectory to trajectory with the aim to fit the gaps between the ground and the first excited states. The  $\tan\chi$ term in (\ref{RM_V1}), equivalently,
the term proportional to $B(R)$ in (\ref{energies_tRM}),  contributes significantly anyway only to the  gaps between the poles characterized  by the lowest $K$ values. {}For $K\geq 2$ values it becomes practically irrelevant and does not affect the linearity of the trajectory at this scale.
The results of the data fits, compiled  in Fig.~\ref{allinone} and the Tables 1-3, are in good agreement with the measurements.

In the Table 2 we list our predictions for the  masses of the 12 ``missing'' mesons discussed above.

In Table 3 we present as an illustrative example of the findings of this work a comparison between the measured and the predicted masses of the mesons on the pion trajectory. The data fits to the mesons on the remaining trajectories are of the same quality and are not given here for the sake of not overloading
the presentation.

As already mentioned to the end of the opening of this section, our model is exclusively  focused on the importance of the external conformal symmetry in shaping the spatial parts of the meson wave functions with the aim to provide an explanation for the observed striking $SO(4)$ degeneracies. The nature of the two meson constituents can only be indirectly deduced from  the properties of their interaction.
To the amount the upgrading interaction used in our analysis is closely related to the Cornell potential, known to describe not only quark-antiquark but also
gluon-gluon ($G-G$) interactions \cite{Buisseret}, we admit for the possibilities that the internal meson configuration  can contain
$(\bar q-q)$ next to $(G-G)$ pairs. 
Within a scheme of such limitations, no prediction on the internal quantum numbers of the mesons, i.e. on their isospins, the $G$ and $C$ parities, can be made. As a guidance in  the assignment of the mesons to the trajectories and their $SO(4)$ poles, we used only the external quantum numbers corresponding to masses, spins, and  spatial parities, which are well defined within the scheme. As additional assumptions, the constancy of the $CP$ parity and isospin over a trajectory has been made. 
A model capable of predicting the internal quantum numbers of the mesons has to be based upon a microscopic Hamiltonian expressed in terms of quark- and anti-quark  creation and annihilation operators, of well defined  properties under charge conjugation and reflection within the isospin space.
Such a model has been developed in Ref.~\cite{Hess} for the particular case of isocalar mesons on the grounds of 
a second quantized Hamiltonian  and with a schematic interaction, meaning that the interaction strength has been maintained as a 
free parameter depending on the quantum numbers characterizing the orbitals and the spins of the quark and anti-quark  creation and annihilation operators.
This model has a total of nine parameters, same as our model.

In order to cross check  consistency of our conjectures on the internal meson quantum numbers with the corresponding definitions in \cite{Hess},
we pick up as a trial set five mesons whose  quantum numbers are equal in both schemes, and compare the predictions on their masses 
by the two respective approaches  (see Table 4). 
In finding congruency between these predictions, we conclude on  the adequacy of  our assumed  $I$ , $G$, and $C$ assignments.\\

\noindent
{}Finally, it needs to be noticed that the potential in (\ref{RM_V1}) 
has been earlier employed by us in the description of baryons, considered as quark-diquark systems, and also there it was shown to provide a very satisfactory data description \cite{CK_2010}, \cite{TQC} not only on the spectra but also of the proton electric charge form factor, though in this reference the very essential  points on the color dipole character of this potential and on its link  to Wilson loops have not yet been understood. We conclude on the relevance of conformally symmetric trigonometric and hyperbolic potentials in quark model physics.

\section{Summary and conclusions}

We developed a unified scheme for the description of me\-sons  as bound states within a well potential, on the one side, and as
resonances transmitted through a barrier, on the other. We employed the pair of potentials constituted by  the trigonometric Scarf well,
$\ell (\ell +1)\sec^2 \chi$, and the P\"oschl-Teller barrier, $\left[(K+1)^2+1/4\right]\sech^2\rho$, where $\ell$ and $K$ stand in turn for the ordinary
and  the four-dimensional angular momentum values. This particular choice of the potential parameters 
guarantees that (i) the real parts of the squared complex resonance energies equal the squared energies of the bound states, 
(ii) the energies are $(K+1)^2$-fold degenerate, thus revealing the conformal symmetry of the spectra.
On these grounds we suggested a conformal symmetry based classification scheme of mesons according to linear resonance trajectories with $SO(4)$ poles
(conformal trajectories), which we  depicted 
on the plane of the total angular momenta ($\ell$) (the mesons integer spins) versus the invariant masses $(M)$ with $\ell$ obeying the $SO(4)$ pole branching rule of, $\ell=0,1,2,..., K$, and for $K=0,1,..,5$.
As examples illustrative of our findings we applied the scheme to the classification of four families of trajectories corresponding to $f_0$,
$a_0$, $\pi$, and $\eta$ meson resonances, on which we could organize into a total of 23 $SO(4)$ poles  71 reported mesons and predict 12 ``missing''.
Only 11 mesons, from them 10 $f$ mesons, dropped out of the systematics.

We observed that with the increase of the energy and especially above $\sim$1400 MeV, the experimental data start aligning better and better to the predicted linearity of the trajectories thus revealing the adequacy of the trigonometric Scarf well in this mass region. As to the mesons of the lowest masses,
the $\sec^2\chi$ interaction failed to predict the correct gaps between the ground state and the first excited states. However, this failure has not been attributed to the violation of the conformal symmetry. The fact is that the extension of the trigonometric Scarf well by the conformally invariant color dipole potential in (\ref{dipole_ctg}), equivalently, the Wilson loop potential in  (\ref{thatis}), was able to satisfactory describe  the aforementioned gaps while continuing respecting the conformal symmetry. In effect, a quite satisfactory nine parameter fit to the 71 classified mesons, distributed over a total of 23 $SO(4)$ poles, could be  performed.
In addition, our analysis showed that above $\sim$1700 MeV the masses of the $SO(4)$ poles from the  respective chiral-partner ($\pi$ - $f_0$), and $(\eta$-$a_0$) 
trajectories start notably approaching each other becoming practically mass degenerate, around $\sim 2000$ MeV and $\sim 2300$ MeV, respectively. We interpreted existence of these trajectories in general and the mass degenerate
parity duplication of the $K=4$ and $K=5$   $SO(4)$ poles in particular,  as a possible signature for chiral symmetry restoration at this mass scale. 
Though an element of  uncertainty is invoked by the status of the resonances placed on the latter poles, predominantly taken from the list of
``Other light mesons '' in \cite{PART},  we nonetheless read this phenomenon as an indication that chiral symmetry for mesons might be realized in terms of parity doubled conformal multiplets rather than, as usually assumed,  in terms of parity doubling of single $SO(3)$ states \cite{Afonin2}.\\

\noindent   
The unified description of bound states and resonances has been achieved by virtue of the four-dimensional de Sitter space time, $dS_4$,
on which the well potential shapes  the free quantum motion on the closed hyper spherical $S^3$ geodesic, while the barrier does same on open hyperbolic geodesics. Further non-trivial insights into the properties of both potentials have been  gained in  formulating effective chromo-statics on
$dS_4$. Namely, it has been found in the Appendix A that the nature of such a statics depends on the type of the $dS_4$ geodesics. Consistent chromo-statics on the unique closed space like geodesic of $dS_4$, the equatorial hypersphere $S^3$, imposes severe restrictions on the allowed number of color charges on this space by  requiring it to be exactly balanced out by an equal  number of anti-color charges. In addition, each single  color-charge potential has to be a fundamental solution of the conformal $S^3$ Laplacian. Then the color--anti-color system creates a color dipole confining potential, shaped after a tangent ($\tan\chi$), or, a cotangent ($\cot\chi$)  function in depending on the range of the argument, $\chi\in[-\pi/2,+\pi/2]$, versus $\chi\in [0,\pi]$, with $\chi$ standing for the second polar angle parametrizing the hypersphere $S^3$. The trigonometric $\sec^2\chi$ function  now takes the part of a color-electric field  to this  potential.
In effect, the $dS_4$ space time provides an intriguing geometric set up, suggestive of
defining a ``geometric  confinement''  as conformal symmetry motivated color neutrality of quark systems placed on $S^3$ closed spaces. This definition, in predicting  conformal  color--anti-color (two-sources)  potentials of the type, $\sec^2\chi +\tan\chi$, equivalently, $\csc^2\chi +\cot\chi$,  could be  tested and convincingly confirmed through a fit to a representative part of the light flavor meson spectrum. It needs to be said that the latter interaction, very well known from the supersymmetric quantum mechanics \cite{susy} under the name of the trigonometric Rosen-Morse potential, has earlier been  used by us also to study baryon spectroscopy in \cite{CK_2010},\cite{TQC}. One of the achievements of the present work relative to the previous ones is to have uncovered its color-dipole nature, visualized in Fig.~8,  
and its origin from cusped Wilson loops in the equation (\ref{thatis}) from Appendix B.   

\noindent
In contrast, on open hyperbolic space times,  such as the $dS_4$ causal patches, which are  Lobachevsky  space time conics of one less 
dimension, ${\mathbf H}_\pm ^3$,  the number of color charges can be arbitrary, and it is within space times of this kind,  where  free color could in principle be released and become observable.  In this way, the de Sitter space time $dS_4$ allows for a geometric expression of confinement. 
Notice that on  ${\mathbf H}_\pm^3$, which does not have  any closed space-like geodesics,   
color neutrality would be possible only on $S^2$ spherical conics, the  quantum motion on which  has to be forced by the color gauge group.
We conclude on the importance for quark models of hadrons of the color-dipole  trigonometric--,  and the free single color hyperbolic potentials, considered in the Appendix B.
{}Finally, again by virtue of the $dS_4$ geometry,  all the involved potentials could be motivated by fundamental principles in 
relating them to Wilson loops with cusps, and to radial quantization. All in all we conclude on the usefulness of the $dS_4$ space time as a tool for modelling the physics of hadrons in support of ref.~\cite{Pereira}, and on the relevance of trigonometric and hyperbolic potentials for constituent quark models.
{To the amount, the $dS_4$ space, in representing  slices of the conformal $AdS_5$ space time, relates to the canonical conformal space time  by a conformal map, physics in both sets of coordinates are equivalent, and the limitations on the color quantum numbers of hadrons imposed by the hyper-spherical geometry of the closed space-like $dS_4$ geodesic (by itself  related to the Minkowski space time by a conformal map)  remain valid also in flat space. Within this context, the $dS_4$ special relativity could be viewed as an aspect of conformal symmetry.   In order to include fermionic hadrons, the color group has to be such that at least two color charges transform as an anti-color, a condition met by $SU(3)_c$. This consideration provides a ``geometric''  argument in favor of the relevance of $SU(3)_C/Z_3$ dynamics of QCD. As long as the only scales which  break beyond doubt  the conformal symmetry are the masses of the QCD quarks, the conformal symmetry is expected to loose power within the heavy flavor sector, an observation supported by the relevance of the conformal symmetry breaking  Cornell potential in this sector. Instead, for the light flavors, we showed that the power Cornell potential needs  an upgrade toward the trigonometric conformally symmetric color dipole potential in (\ref{dipole_ctg_1}). For a conformally not symmetric strong dynamics the limitations on color neutrality will no longer be compelling and, in case the conformal symmetry were to be the only reason behind the color neutrality,  one may entertain the possibility of observing  heavy flavor free color charges. At any rate, in our opinion, conformal symmetry breaking, signaled  by possibly  significant deviations of the heavy flavor meson spectra from those of the light flavor mesons,
could turn out to be a precondition for deconfinement.}   

Without entering into details, the $dS_4$ geometry, also reflects, at the quantum mechanical level, the fundamental relevance of $AdS_5/CFT_4$ for  QCD at high excitations.

\noindent
{\bf Acknowledgments}\\
We express our gratitude to Dr. Abdulaziz Alhaidari for a detailed illuminating correspondence  on the properties of the Manning-Rosen barrier.

\section*{Appendix A: Chromo-statics on $S^3\in dS_4$.  The sec$^2\chi $ function as  an ${\mathbf E}$  field of a conformally symmetric  color  dipole confining po\-ten\-tial shap\-ed after a tangent function }

The $dS_4$ space-time has a rich geometric structure, as visible from  the Figure \ref{Hyperborea}. The four dimensional hyperboloid of one shell, embedded in a five dimensional Minkowski space time,  can be time-sliced  on closed, open or flat subspaces. The closed slices are ${\mathcal R}^1\otimes S^3$ conformal compact space times, the open ones are three dimensional two-sheeted hyperbolic subspaces, here denoted by ${\mathcal R}^1\otimes {\mathbf H}_\pm^{3}$, and the flat slices are just $M^{2,1}$  planes.  We here are especially interested in the $dS_4$ equator, which is the three sphere, $S^3$, at $\rho=0$. 
The hyperbolic angle encodes the size of the $S^3$ radius away from the equator, and therefore, as it will be shown in the next section, a time variable, meaning that we here are addressing the static case.\\

\noindent
To be specific, we are interested in the possibility to formulate effective (one-color) chromo-statics on $S^3$. In parallel to electrostatics, one may start approaching this goal by considering the fundamental solutions to the $S^3$ Laplace operator in (\ref{LB_S3}). It is not difficult to cross check that the   $(-\frac{q}{4\pi}\tan \chi)$ function, where $q$ stands for a color charge, is such a solution,
\begin{equation}
\Delta_{S^3}(\chi,\theta,\varphi)\tan\chi =0.
\label{crvd_CLMB}
\end{equation}
The tangent function parallels on $S^3$ the $(-1/r)$ fundamental solution to the three dimensional flat space Laplacian, $\nabla^2$, 
and could be thought of  as a version of a  ``curved Coulomb'' potential \cite{BarutWilson}. The corresponding  ${\mathbf E}(\chi)$ field is then obtained as,  
\begin{equation}
{\mathbf E}(\chi) = -\frac{q}{4\pi } \frac{\partial (-\tan \chi )}{\partial \chi}= \frac{q}{4\pi \cos^2\chi}=\frac{q}{4\pi}\sec^2 \chi.
\label{grad_pot}
\end{equation}
Therefore, the $\sec^2\chi$ function   is found to shape  on $S^3$ a  static color-electric  field, denoted by ${\mathbf E}(\chi)$. 
A similar result, though for electric charges, has been reported in \cite{PouriaPedram}, where it has been derived from the Gau\ss\, theorem on $S^3$.  
However, along this line of reasoning one encounters a serious problem. Suppose, for concreteness, that the charge has been located at $\left[\chi=-\pi/2, \varphi=-\pi/2\right]$, referred to here as the ``West'' pole.
Because the outgoing field lines are confined to remain on the hypersphere, they will follow the great circles, which necessarily intersect for a second time as ingoing lines at $\left[ \chi=\pi/2, \varphi=\pi/2\right]$, i.e. at the precise anti-pod, the  ``East'' pole, thus creating there a fictitious source whose charge 
is opposite to the physical one in (\ref{grad_pot}). This is of course an unacceptable unphysical situation in several aspects, one of them,
also noticed in \cite{PouriaPedram}{}, concerns the violation of the superposition principle on $S^3$. We here solve this dilemma by noticing that 
this inconsistency appears in consequence of first setting $\rho$ (equivalently, the time) to a constant, and then considering $S^3$, instead of first considering
the full conformal ${\mathcal R}^1\otimes S^3$ space time and then setting $\rho$  to a constant. As it will be shown in section 5.1 below, in the latter case   
one encounters the conformal Laplacian, here denoted by, $-\Delta^1_{S^3}(\chi,\theta,\varphi)$,   and presented (in dimensionless units) below in the equation (\ref{cnfrm_LPL}) as,
\begin{eqnarray}
-\Delta_{S^3}^1(\chi,\theta,\varphi)&=&{\mathcal K}^2(\chi,\theta,\varphi) +1.
\label{CNFRM_LPL}
\end{eqnarray}
The fundamental solutions (Green functions) ${\mathcal G}_{-\frac{\pi}{2}}(\chi)$, and ${\mathcal G}_{\frac{\pi}{2}}(\chi)$  to the conformal Laplacian on the hypersphere, corresponding to color-charges placed at $\chi=-\pi/2$, and 
$\chi=+\pi/2$ (``West'' and ``East'' poles, respectively) are now distinct. They  have been calculated for $R=1$ for example  in \cite{Stephani}{}, \cite{Alertz} as
\footnote{In these references the parametrization of the sphere is such that the  second polar angle varies as $\chi\in [0,\pi]$, while we here use instead
$\chi\in [-\pi/2,+\pi/2]$,  a reason for which  our $\tan\chi$ changes  to their $\cot\chi$. }
\begin{eqnarray}
{\mathcal G}_{-\frac{\pi}{2}}(\chi) &=&\frac{1}{4\pi^2}
\left(\frac{3\pi}{2} -\chi\right) \tan \chi +c_0,\label{GNorth}\\ 
{\mathcal G}_{+\frac{\pi}{2}}(\chi)  &=& \frac{1}{4\pi^2 }
\left(\frac{\pi}{2}- \chi\right) \tan \chi +c_1,
\label{GSouth}
\end{eqnarray} 
where $c_0$ and $c_1 $ are  constants. 
The \underline{single color}  potentials associated with  these Green functions are,
\begin{eqnarray}
{\mathcal V}_{-\frac{\pi}{2}}(\chi)&=&q_1{\mathcal G}_{-\frac{\pi}{2}}(\chi) =\frac{q_1}{4\pi^2}
\left(\frac{3\pi}{2} -\chi\right) \tan \chi +q_1c_0,\nonumber\\
 q_1 &=&-q,\label{GNorth_pot}\\ 
{\mathcal V}_{+\frac{\pi}{2}}(\chi)&=&q_2{\mathcal G}_{+\frac{\pi}{2}}(\chi)  = \frac{q_2}{4\pi^2 }
\left(\frac{\pi}{2}- \chi\right) \tan \chi +q_2c_1,\nonumber\\
 q_2&=&q,
\label{GSouth_pot}
\end{eqnarray} 
respectively.
The tangent function in (\ref{grad_pot}) can be recovered but now as a dipole potential produced by two real physical  pod-anti-pod charges 
and expresses in terms of the fundamental solution of the conformal Laplacian in (\ref{GNorth})-(\ref{GSouth}) as,
\begin{eqnarray}
{\mathcal V}_{-\frac{\pi}{2}}(\chi)+{\mathcal V}_{+\frac{\pi}{2}}(\chi)=
q_1{\mathcal G}_{-\frac{\pi}{2}}(\chi) +q_2{\mathcal G}_{+\frac{\pi}{2}}(\chi)& =&-\frac{q}{4\pi}\tan \chi,\nonumber\\
 c_1 &=&c_0.
\label{dipole_ctg}
\end{eqnarray}
In effect, the  ${\mathcal V}_{-\frac{\pi}{2}}(\chi) $ and ${\mathcal V}_{+\frac{\pi}{2}}(\chi) $ functions describe
two different potentials, the first  generated by a physical single color charge-, and the second--by  a physical  anti-color charge, while the tangent function appears as the 
associated color dipole potential. In this way, the inevitable  $S^3$ charge neutrality is respected  but it stops being unphysical.
Now the fields of the physical ``negative'' ($q_1=-q$), and ``positive'' ($q_2=q$) color charges  related to the potentials in the equations (\ref{GNorth_pot}) and (\ref{GSouth_pot}) are easily calculated and read,

\begin{eqnarray}
{\mathbf E}_{-\frac{\pi}{2}}(\chi) &=&
-\frac{\partial }{\partial \chi}\frac{q_1}{4\pi^2}\left( \frac{3\pi }{2}-\chi \right)\tan\chi =\frac{q}{4\pi^2}\left(\frac{3\pi }{2}-\chi  \right)
\nonumber\\
&\times& \frac{1}{\cos^2\chi}-\frac{q}{4\pi^2}\tan \chi, \nonumber\\
{\mathbf E}_{+\frac{\pi}{2}}(\chi) &=&
-\frac{\partial }{\partial \chi}\frac{q_2}{4\pi^2}\left( \frac{\pi }{2}-\chi \right)\tan\chi =-\frac{q}{4\pi^2}\left(\frac{\pi }{2}-\chi  \right)
\nonumber\\
&\times& \frac{1}{\cos^2\chi}
+\frac{q}{4\pi^2}\tan \chi.
\label{rphys_flds}
\end{eqnarray}
Their superposition is now well defined and amounts to,
\begin{equation}
{\mathbf E}_{+\frac{\pi}{2}}(\chi) +{\mathbf E}_{-\frac{\pi}{2}}(\chi) =\frac{q}{4\pi \cos^2\chi}={\mathbf E}(\chi),
\label{lin_sprp}
\end{equation}
which formally  coincides with the chromo-static field  in (\ref{grad_pot}). 
However, now the $\sec^2\chi $ function has been found as the shape of an ${\mathbf E}$  field due to a charge  dipole-, and not to a single charge potential,
as supposed in (\ref{grad_pot}) and \cite{PouriaPedram}. Therefore, the color-electric field on $S^3$ takes  the shape of the $\sec^2\chi$ function, while the
dipole potential is generated by a colorless two-body system, like mesons,  and in agreement with the inevitable  $S^3$ charge neutrality.

\begin{quote}
Stated differently, on $S^3$, which we introduced as the closed space-like geodesic on $dS_4$,
there can be only color-anti-color charge pairs generating conformal color dipole-potentials, shaped after a tangent function,
and an associated conformal ${\mathbf E}$ field, shaped after the $\sec^2\chi $ function. In this manner, this manifold
\begin{itemize}  
\item provides a geometric set up suited for modelling  confinement within the environmental space as color neutrality of strong interacting systems,
\item links this confinement to conformal symmetry of closed space times.
\end{itemize}

\end{quote}
In contrast, on the $dS_4$  open hyperbolic geodesics (they are $dS_3$ spaces), and/or on the  open conic sections, any arbitrary number of color-neutral  and single-color systems are allowed to propagate  independently from each other.
Single color (single source) potentials on $dS_3$ are associated with the Green function of the $dS_3$  conformal Laplacian (in dimensionless units), 
\begin{eqnarray}
-\Delta_{dS_3}^1(\rho,\theta , \varphi) &=&-\Delta_{dS_3} (\rho, \theta,\varphi) +1,
\end{eqnarray}    
in which case one encounters in the hyperbolic variable the free color-charge potential as $\left( -\frac{q}{4\pi^2 }\rho \tanh \rho \right) $. 
Stated differently, open space times do not require the number of color charges to equal the number of anti-color charges. 

\noindent
As a final remark, we wish to recall that the phenomena considered so far,  are all described  by space-like physics. Indeed, the well-potential represents an instantaneous interaction, and the barrier potential describes, via tunneling,  virtual resonance transmission in scattering. The space-like $dS_4$ surface provided the appropriate set-up for their description and interrelation. We conclude on the relevance of $dS_4$ symmetry and therefore on $dS_4$ special relativity for QCD processes in the space-like region, in support of ref.~\cite{Pereira}. Off-shell strong-interactions seem to be sensitive to higher dimensions.\\

\noindent
All these clues could be gained thanks to appropriate variable changes and coordinate transformations 
in the quantum mechanical wave equations, and  in combination with convenient choices for the magnitudes of the  potentials under investigation.\\

\noindent 
We also like to remark that exploring a comment by Love\-lace \cite{Lovelace},  we first tried to depart from  the spectrum of the $\ell (\ell +1)\csc^2\chi$ potential and to transform it into the
Manning-Rosen barrier,
$\left[(K+1)^2+1/4\right]\csch^2\rho$, in which case we  approached instead of $dS_4$,  the causal four-di\-men\-si\-onal hyperboloid of two sheets,
${\mathbf H}_\pm ^4$, also referred to as Lo\-ba\-chev\-sky's four plane. However, physics with the Manning-Rosen barrier is quite distinct from that of the P\"oschl-Teller barrier. The Manning-Rosen barrier, singular at origin, does not allow for transmission \cite{Gadella}{},\cite{Ahaidari}{},\cite{MattVisser}  and no duality can be constructed  between the bound states within the $\csc^2\chi $ well (identical to those of the $\sec^2\chi$ well) and states residing  in Lobachevsky's hyperbolic space time. On the latter, no confinement can be defined in terms of quantum motion along geodesics, at most, it could  be defined as  motion on closed space like conics.  Free color charges on ${\mathbf H}_\pm ^4$ will give rise to  $\left( -\frac{q}{4\pi^2 }\eta \coth\eta \right)$ potentials, where $\eta$ is now the hyperbolic angle on the respective Lobachevsky space-time.  \\

\noindent
The above considerations reveal the unifying r\'ole played by large  geometric set ups with respect to  seemingly disconnected processes in flat space. On 
 curved surfaces, the distinct processes  can acquire meaning of different aspects of a more general process.

In this fashion,  multidimensional hyperbolic set ups  could be employed as useful tools in modelling  at the quantum mechanical  level of complex physical phenomena.

\subsection{Conformal extension of  the trigonometric Scarf  well  by the confining color  dipole potential }

The dipole potential introduced  in (\ref{dipole_ctg})  in terms of the Green function of  the conformal Laplacian, $\left({\mathcal K}^2(\chi,\theta,\varphi)+1\right)$, has interesting properties with respect to ${\mathcal K}^2(\chi,\theta,\varphi)$.  It commutes with the latter,  and acts as an isometry,
\begin{equation}
{\mathcal K}^2(\chi,\theta,\varphi) \tan\chi\,  f(\chi,\theta,\varphi)=\tan\chi {\mathcal K}^2(\chi,\theta,\varphi)  \, f(\chi,\theta,\varphi),
\label{isometry}
\end{equation} 
where $f(\chi, \theta,\varphi)$ is any arbitrary differentiable function. In other words, the color dipole potential is necessarily $SO(4)$ symmetric.
One can now use it as a perturbation of the kinetic quantum motion on $S^3$ and write down the following wave equation:

\begin{eqnarray}
{\mathcal H}(\chi)\widetilde{\psi}^{(b)}_{\ell n }(\chi )
=\left(\epsilon^{(b)}_{\ell n}\right)^2  \widetilde{\psi}^{(b)}_{\ell n }(\chi ),&&\nonumber\\
{\mathcal H}(\chi) =-\frac{1}{\cos^2 \chi}\frac{\partial }{\partial \chi }\cos^2 \chi \frac{\partial }{\partial \chi }
+\frac{\ell(\ell +1)}{\cos ^2\chi } -2b\tan \chi +1.&&\nonumber\\
\label{RosMor}
\end{eqnarray}
Upon the variable change, 
\begin{equation}
\widetilde{\psi}^{(b)}_{\ell n }(\chi )\cos\chi = U_{\ell n}^{(b)}(\chi),
\label{var_chng_5}
\end{equation}
the  equation (\ref{RosMor}) takes the shape of an one-dimensional wave equation with the so called trigonometric Rosen-Morse potential:
\begin{eqnarray}
\left( -\frac{{\mathrm d}^2}{{\mathrm d}\chi^2} +V^{(b)}_{\mbox {tRM}}(\chi)\right)
 U_{\ell n}^{(b)}(\chi)=\left({\epsilon}_{\ell n}^{(b)}\right)^2 U_{\ell n}^{(b)}(\chi),
&&\nonumber\\
V^{(b)}_{\mbox {tRM}}(\chi)=\frac{\ell(\ell +1) }{\cos^2\chi} -2b\tan \chi,&&\nonumber\\
 \left({\epsilon}_{\ell n}^{(b)}\right)^2=-\frac{b^2}{\left( K+1 \right)^2} +(K+1)^2,&&\nonumber\\
 K=\ell +n,\quad \chi \in \left[-\frac{\pi}{2},+\frac{\pi}{2} \right],&&\nonumber\\
\label{RM_V1}
\end{eqnarray}
where we set $\hbar=1$, $c=1$, and $R=1$,  thus writing down the wave equation in the widely used dimensionless units \cite{susy}.
In this units, and for $b=0$,  the energy $ \left({\epsilon}_{\ell n}^{(b)}\right)^2$ recovers $\left( {\mathcal E}^{(bound)}_{2} \right)^2 R^2/(\hbar^2c^2)$ 
in the equation (\ref{ScarfI}). For $b\not=0$ the $(K+1)^2$-fold conformal degeneracy is still respected
meaning that the extended potential continues describing  a conformally invariant confined  colorless system.
Therefore, the conformal symmetry on the closed space  is fully  respected by the color dipole interaction.

Thanks to the negative sign of the $b$ term, the $V^{(b)}_{\mbox {tRM}}(\chi)$ potential provides an efficient tool for increasing the splitting between the ground and lowest excited states in 
the $\sec^2\chi$ well and thus for improving description of meson data, an option of which we make use in the subsequent section.
On $dS_4$ this potential is converted into the hyperbolic Rosen-Morse potential \cite{susy}{},

\begin{eqnarray}
\left( -\frac{{\mathrm d}^2}{{\mathrm d}\rho^2} +V^{(b)}_{\mbox{hRM}}(\rho)\right)
 U_{\ell n}^{(b)}(\rho )=\left({\epsilon}_{\ell n}^{(b)}\right)^2 U_{\ell n}^{(b)}(\rho ),&&
\nonumber\\
 \left({\epsilon}_{\ell n}^{(b)}\right)^2=-\frac{b^2}{\left( K+1 \right)^2} -(K+1)^2,&&\nonumber\\
K=\ell -n, \quad \rho \in \left[-\infty,+\infty  \right],&&\nonumber\\
V^{(b)}_{\mbox{hRM}}(\rho)=-\frac{\ell(\ell +1) }{\cosh^2\rho } -2b\tanh \rho.&&
\label{RM_HPRB}
\end{eqnarray}
To the amount, the $\sech^2\rho$ and $\tanh\rho$  potentials emerge from the analytical continuation of the $\sec^2\chi$ and $\tan\chi$ potential towards complex values of the argument (in combination with $b\to -ib$), we here take the position that also the hyperbolic Rosen-Morse potential describes quantum motion of colorless systems though such  propagating off-shell along open hyperbolic geodesics on $dS_4$.  In contrast, the free color-charge potential is expected to be shaped after, $\left( -\chi\tanh\chi\right)$.

However, the more familiar notations for the trigonometric Rosen-Morse potentials are those in which the second polar angle, $\chi$,  on $S^3$ varies
as $\chi\in \left[0,\pi\right]$.
In this parametrization of $S^3$ one finds,
\begin{eqnarray}
\left( -\frac{{\mathrm d}^2}{{\mathrm d}\chi^2} +\frac{\ell(\ell +1) }{\sin^2\chi} -2b\cot \chi\right)
 R_{\ell n}^{(b)}(\chi)&=&\left({ \epsilon}_{\ell n}^{(b)}\right)^2 R_{\ell n}^{(b)}(\chi),\nonumber\\
\chi &\in& \left[0,\pi \right].
\label{RM_V2}
\end{eqnarray}
In this coordinate choice the Green functions in (\ref{GNorth}) and (\ref{GSouth}) change correspondingly to,
\begin{eqnarray}
G_{-\frac{\pi}{2}}(\chi)\to G_{ 0}(\chi) =\frac{1}{4\pi^2}(\pi -\chi)\cot \chi +c_0,
\label{GNorth_1}\\
G_{\frac{\pi}{2}}(\chi) \to G_{ \pi }(\chi) =-\frac{1}{4\pi^2}\chi \cot \chi +c_1,
\label{GSouth_1}
\end{eqnarray}
while the dipole potential in (\ref{dipole_ctg}) becomes,
\begin{equation}
q_1{\mathcal G}_{o}(\chi)  +q_2{\mathcal G}_\pi (\chi)  =-\frac{q}{4\pi}\cot \chi , \quad c_1=c_0, \quad q_1=-q_2=-q.
\label{dipole_ctg_1}
\end{equation}
The associated ${\mathbf E}$ field is then given by a $\csc^2\chi $ instead of a $\sec^2\chi $ function,
\begin{eqnarray}
{\mathbf E}(\chi)&=&\frac{q}{4\pi}\frac{1}{\sin^2\chi}.
\label{ENEW}
\end{eqnarray}
In the light of the results presented at the beginning of the current section, the reading which  we give to the  color dipole potential, 
 be it in the parametrization of (\ref{dipole_ctg}) or (\ref{dipole_ctg_1}), is that
{\it  conformal symmetry on closed surfaces  is closely related to confinement.\/}
In the following we shall switch  to the  notation in (\ref{dipole_ctg_1}) for the purpose of staying closer to the nomenclature used in the current literature \cite{susy}. The reason for which so far we have been working with the alternative $\chi\in [-\pi/2,+\pi/2]$ parametrization of $S^3$ was to circumvent that the analytical continuation of the well potential amounted to the singular csch$^2\rho $ Manning-Rosen barrier, discussed to the end of the previous subsection.
We hope that this new nomenclature has been made clear and that no confusion will arise in the following.

The parametrization of the color dipole potential in (\ref{RM_V2}) has the advantage that it visualizes better the four important properties of the effective  quark potential, namely, the
\begin{itemize}
\item color-neutrality of quark systems,
\item  asymptotic freedom,
\item anti-screening,
\item relevance of the also two-sources Cornell potential predicted by lattice QCD \cite{Cornell}.
\end{itemize}
All these properties are visualized in Figure~3.
\begin{figure}
\resizebox{0.43\textwidth}{4.5cm}
{\includegraphics{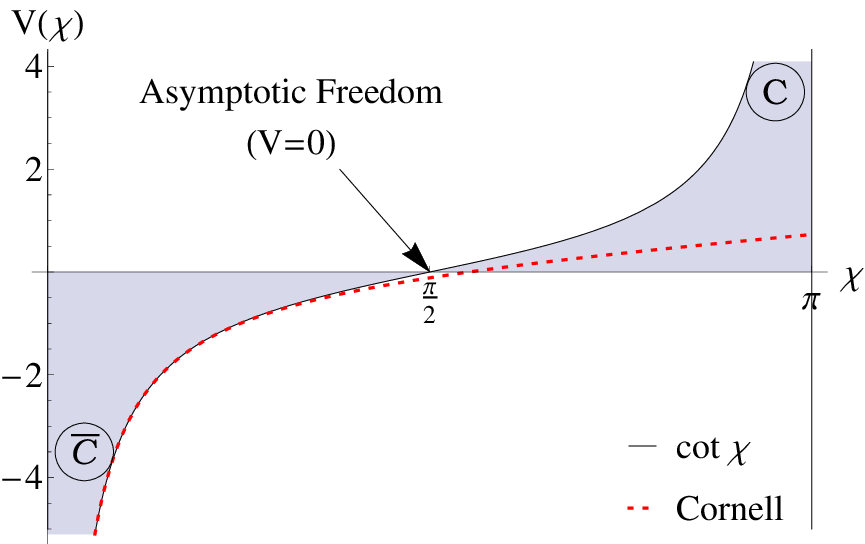}}
\caption{ The color dipole potential in the equations (\ref{dipole_ctg_1})-(\ref{Cornel}) and its characteristics. Charge and anti-charge have been denoted by $C$ and ${\overline C}$, respectively.
\label{confpt}}
\end{figure}
The dipole character of the potential, becomes apparent on the graphical display through the  
$-\infty$, and $+\infty$-divergences brought about by the two sources located at $\chi\to 0$, and $\chi\to \pi$, respectively.
As to the asymptotic freedom, it is evident through the fact that as $\chi\to \pi/2$, where  the potential becomes vanishing, the quarks trapped by the cotangent potential
will behave as asymptotically free. In contrast, at larger distances, $\chi \to \pi$, the potential grows infinitely thus evolving to anti-screening. 
{}Finally, the first terms in  the Taylor series expansion of $\cot\chi$ feature the inverse distance plus linear  potential
predicted by lattice QCD according to \cite{TQC},
\begin{equation}
-2b \cot\chi=-\frac{2b}{\chi }+\frac{4b}{3}\chi +{\mathcal O}(\chi^2), \quad \chi =\frac{\stackrel{\frown}{r}}{R},
\label{Cornel}
\end{equation}
where $ \stackrel{\frown}{r}$ is the arc distance of the anti-quark from the quark located  at the  North pole (or, vice versa). 
The ${\mathcal O}(\chi^2)$ terms  in (\ref{Cornel}) can be viewed as a phenomenological corrections to the Cornell potential which serves as an interaction between massless gluons besides between a quark and an anti-quark \cite{Buisseret}.
In the next section we show how  the potentials discussed so far relate to Wilson loops with cusps.

\section*{Appendix B: Origin of the conformally symmetric  color dipole potential  from  cusped  Wilson loops and conformal symmetry motivated definition of a  geometric confinement as color neutrality on closed spaces }

Wilson loops with cusps are directly related to  quark phenomenology \cite{Belitsky}. Indeed, at the discontinuity in the velocity function, referred to as a ``cusp'', the quark of mass $m$ promptly changes its direction of propagation
from velocity  $v_\mu$ to $v_\mu^\prime$  with $<\!\!\!) v.v^\prime=\chi  $, where
the angle $\chi$ has been chosen in such a way that $\chi=\pi$ lies on  the straight line from which the deviation is measured (for example,  the direction of $v_\mu$).
Due to the sudden acceleration the quark starts emitting soft 
(virtual and real) gluons with momenta $k<\mu$ where $\mu$ is  a cut-off of the order of the quark mass. It is commonly accepted 
to treat the soft gluon emission from fast quarks for both the incoming and outgoing quarks in the eikonal approximation, 
in which  the color-charged particles are supposed to behave classically. Then, their interactions with the soft gluons can be described 
by path ordered Wilson lines along their classical trajectories \cite{Gorsky_IK}. In due course,
a static potential is generated,  whose source is, say,  a quark $(q)$ and which can be perceived by a nearby located anti-quark $(\bar q)$.
Such potentials are  worth  being explored in spectroscopic studies.
In particular, the  expectation value of a Wilson loop over a rectangular path of height $T$ and width $L$ relates to a potential, $V(L)$ 
according to \cite{Zlotnikov}
\begin{equation}
<W (v\cdot v^\prime) >= <P\exp i\oint  {\mathrm x}_\mu A^\mu (x)>=\exp \left( -TV(L) \right).
\label{wilson1}
\end{equation}
For $L<<T$  the Wilson loop vacuum expectation value encodes the interaction between two ``trajectories'', 
one of which is quarkish, and the other- anti-quarkish, 
\begin{eqnarray}
<W(v\cdot{ v^\prime}) >&=&\exp \left( -TV (L) \right), \quad L<< T.
\label{wilson2}
\end{eqnarray}
{}For one-cusp integration contours, one finds \cite{Gorsky_IK}
\begin{eqnarray}
<W(v\cdot v^\prime) >&=&1- \frac{\alpha_s N_c}{\pi}\left(w(v\cdot v^\prime) -w(1)\right) +{\mathcal O}(\alpha_s^2),\nonumber\\
\label{wilson3}
\end{eqnarray}
where $\alpha_s$ is the strong coupling, $N_c$ stands for  the number of colors, and $w(v\cdot v)=w(v^\prime \cdot v^\prime)=w(1)$.
The $w(v\cdot v^\prime) $ function depends on the 
angle  $\chi$  between the two velocities $v_\mu$ and $v_\mu^\prime$, at the cusp, and will be from now onward re-denoted as
\begin{equation}
w(v\cdot v^\prime )\longrightarrow w(\chi).  
\label{wfnct_rlbld}
\end{equation}
Along the line of Ref.~\cite{Gorsky_IK} one calculates,
\begin{eqnarray}
w(\chi  )=
\int_{-\infty} ^0 {\mathrm d}s \int_0^\infty {\mathrm d}t \frac{v\cdot v^\prime}{(vs-v^\prime t)^2}
&=& \Gamma_{\mbox{cusp}}(\chi  ) \ln \frac{L_{IR}}{\epsilon_{UV}},\nonumber\\
 \cos\chi &=&\frac{v\cdot v^\prime}{\sqrt{v^2}\sqrt{v^\prime{}^2}},
\label{cuspintegral}
\end{eqnarray}
where $s$ and $t$ have the meaning of proper times. The 
cusp function, $\Gamma_{\mbox{cusp}}(\chi  )$, obtained in this way reads,
\begin{eqnarray}
\Gamma_{\mbox{cusp}}(\chi  )&=&\chi \cot \chi +{\bar c}_1,
\label{cuspfunction}
\end{eqnarray}
where ${\bar c}_1$ is a constant. Furthermore,  $L_{IR}$ and $\epsilon_{UV}$ are the  cut offs  in the Infrared and Ultraviolet regimes of QCD,
respectively. Therefore, a cusped Wilson loop is logarithmically divergent according to \cite{Gorsky_IK},
\begin{equation}
<W>_{\mbox{cusp}}\sim e^{-\frac{\alpha_s  N_c}{\pi }\Gamma_{\mbox{cusp}}(\chi) \ln \frac{L_{IR}}{\epsilon_{UV}}}.
\label{logdiv}
\end{equation}

On the way of  finding an interpretation for the potential in (\ref{wilson2}) associated with the cusp function in (\ref{cuspfunction}),
 one  starts with a four-dimensional Euclidean space,
$E_4$, (also referred to as 4 plane, ${\mathcal R}^4$, or, quaternion plane) parametrized in  Cartesian 
 coordinates, $x_4$ and $\vec{x}$. This space represents a comfortable testing ground for theoretical considerations in so far as it relates to the physical relativistic   Minkowski space by a  Wick rotation, $x_4\to ix_4$. 
The ${\mathcal R}^4$ plane is  parametrized  by real quaternions as  \cite{Fubini},\cite{Kirtisis},
\begin{eqnarray}
{\mathcal R}^4:\quad Z&=& x_4+i{\vec \sigma}\cdot {\vec r},\quad {\vec r}=(x_1,x_2,x_3),\nonumber\\
{\vec \sigma}&=&(\sigma_1,\sigma_2,\sigma_3),
\label{E4_plane}
\end{eqnarray}
where $\sigma_i$ are the Pauli matrices, while $i\sigma_i=e_i$ are Hamilton's  three imaginary units.  In polar representation, the quaternion is given by,
\begin{eqnarray}
Z=R e ^{i \chi A} =R\cos \chi\,  {\mathbf 1_{2\times 2}}  +iAR\sin\chi,&&\nonumber\\
\chi\in \left[0,\pi\right],&&\nonumber\\
{\bar Z}= R e ^{-i \chi A} =R\cos \chi\, {\mathbf 1_{2\times 2}}  -iAR \sin\chi,&& \nonumber\\
 R^2=x_4^2+x_1^2+x_2^2+x_3^2=e^{2\tau}, &&\nonumber\\
A=\left(
\begin{array}{cc}
\cos \theta&\sin\theta e^{-i\varphi}\\
\sin\theta e^{i\varphi}&-\cos\theta
\end{array}
\right), \quad \theta\in \left[ 0,\pi \right],&&
\label{trd_prmtrz}
\end{eqnarray}
and $\quad A^2={\mathbf 1_{2\times 2}}$. Here $\theta$ and $\chi$ are in turn the first and second polar angles in ${\mathcal R}^4$, $\varphi$ is the azimuthal angle there, while $\tau=\ln R$ is the so called conformal time.
Alternatively, for $\chi\to \left( \chi -\pi/2\right)$, the  conformal map in (\ref{trd_prmtrz}) changes shape to
\begin{eqnarray}
Z=R e ^{i \chi A} =R\sin \chi\,  {\mathbf 1_{2\times 2}}  +iAR\cos\chi, &&\nonumber\\
 \chi \in \left[-\frac{\pi}{2},+\frac{\pi}{2} \right], &&\label{gleich1}\\
{\bar Z}= R e ^{-i \chi A} =R\sin \chi\, {\mathbf 1_{2\times 2}}  -iAR \cos\chi, &&\label{gleich2}\\
 R^2=x_4^2+x_1^2+x_2^2+x_3^2=e^{2\tau}, && \nonumber\\
A=\left( \begin{array}{cc}
\sin \theta&\cos \theta e^{i\varphi}\\
\cos \theta e^{-i\varphi}&-\sin\theta
\end{array}
\right), \quad \theta\in \left[ -\frac{\pi}{2},+\frac{\pi}{2}\right].&&
\label{gleich3}
\end{eqnarray}

In continuation, the parametrization in (\ref{trd_prmtrz})  will be predominately used for the sake of respecting preferences in the literature on the subject. Then the map, 
\begin{eqnarray}
w=\tau +i A\chi=\ln Z=\ln (x_4+i{\vec \sigma}\cdot {\vec r}),
\end{eqnarray}
establishes a correspondence between  the four plane  and the  ${\mathcal R}^1\otimes S^3$ cylinder,
\begin{equation}
{\mathcal R}^4\leftrightarrow {\mathcal R}^1\otimes S^3.
\label{map_flat_curved}
\end{equation}
Stated differently,  the three spatial dimensions  , $x_i$, for $i=1,2,3$, have been compactified, $x_i=x_i+2\pi$
( for the purpose of avoiding infrared divergences \cite{Kirtisis}) thus ending up with  ${\mathcal R}^1\otimes S^3$, also  known under the name of a compactified Minkowski space time \cite{LuscherMack}. The conformal map takes the flat-space Laplace operator,
\begin{eqnarray}
\Delta_{M^{3,1}} (x_4,x_1,x_2,x_3)& =& \frac{\partial ^2}{\partial x_4^2}+ \frac{\partial ^2}{\partial x_1^2}+ \frac{\partial ^2}{\partial x_2^2}
+\frac{\partial ^2}{\partial x_3^2},
\label{E4_LB}
\end{eqnarray}
to
\begin{eqnarray}
\Delta_{M^{3,1}}(R,\chi,\theta,\varphi)  & =& \frac{1}{R^3}\frac{\partial }{\partial R}R^3 \frac{\partial }{\partial R}
-\frac{1}{R^2}{\mathcal K}^2(\chi,\theta,\varphi),\nonumber\\
\label{E4_LB_polar}
\end{eqnarray}
with ${\mathcal K}^2(\chi, \theta,\varphi)$ being already defined in (\ref{LB_S3}).
Its eigenvalue problem reads, 
\begin{eqnarray}
{\mathcal K}^2 (\chi,\theta,\varphi)  \Psi(R,\chi,\theta,\varphi) &=&K(K+2)\Psi (R,\chi,\theta,\varphi),\nonumber\\
\label{LB_EVP}
\end{eqnarray}
where $K$ is the value of the angular momentum in four Euclidean dimensions, also  
previously defined in (\ref{quantumnumbrs}), (\ref{PSDAM_DEF}).
Introducing the new variable, $\tau$, as
\begin{equation}
R(\tau)=e^{\tau}, \quad \mbox{with}\quad {\mathrm d}R(\tau)=R(\tau){\mathrm d}\tau, 
\label{cnfrm_tme}
\end{equation}
renaming   $ \Psi(R,\chi,\theta,\varphi)$ by,
\begin{eqnarray}
\Psi (R(\tau),\chi,\theta,\varphi)= U(\tau, \chi,\theta,\varphi),
\label{wavu_var_chng}
\end{eqnarray}
and substituting in (\ref{E4_LB_polar}) gives,
\begin{eqnarray}
e^{2\tau}\Delta_{M^{3,1}}^1 (\tau,\chi,\theta,\varphi)U(\tau,\chi,\theta,\varphi) &=&
\left[  \frac{\partial^2}{\partial \tau ^2} -\left({\mathcal K}^2(\chi,\theta,\varphi)+1\right) \right]\nonumber\\
&\times&U(\tau,\chi,\theta,\varphi).
\label{CLB_step1}
\end{eqnarray}
The latter equation shows that  $\tau$ plays the  r\'ole of a time variable, termed to as ``conformal time'', and that one has switched from the
four dimensional Euclidean plane in (\ref{E4_plane})  to the cylindrical space-time, ${\mathcal R}^1\otimes S^3$ in
(\ref{map_flat_curved})  with the hyper spherical spatial geometry.
In effect, equal  times on the cylinder become $S^3$ spheres of equal radii. In consequence,  time orderings are replaced by
``radial orderings'', and the above framework has become  known as ``radial quantization''\cite{Fubini}.
The infinite past and infinite future map correspond in their turn to a  vanishing, and an infinite  radius.
{}For constant times, the conformal Laplacian on the hypersphere,  denoted by, $\Delta^1_{S^3}(\chi,\theta,\varphi)$, is just,

\begin{equation}
-\Delta^1_{S^3}(\chi,\theta,\varphi)={\mathcal K}^2 (\chi,\theta,\varphi)+1, \quad \tau =\mbox{const}.
\label{cnfrm_LPL}
\end{equation}
This is precisely the correct Laplacian which we employed  in (\ref{CNFRM_LPL}) in the construction of chromo-statics on $S^3$. 
To the amount this statics turned out to refer to colorless systems, the  conformal symmetry has been linked to confinement in the way discussed after the equation (\ref{ENEW}).\\

\noindent
{}Furthermore, the generator of dilatation on the plane, $Z=\lambda Z$, has become translation by $(\ln \lambda)$ in the radial direction 
of the cylinder as it sums up with $\tau$ in the conformal map, a reason for which the dilatation operator on the plane corresponds 
to the kinetic energy on the sphere, and thereby to the $\Delta_{M^{3,1}}^1(\tau,\chi,\theta,\varphi)$ operator in (\ref{CLB_step1}).
If the theory is conformally invariant, the two descriptions a completely equivalent \cite{Kirtisis}{}, \cite{Lovelace}{}.
The circumstance that the dilation operator on the plane becomes the Hamiltonian on the hypersphere provides an alternative view on  the 
cusp function. The issue is that at constant times (constant radius, equivalently, constant $\rho$) the eigenvalues of the conformal Laplacian operator $\Delta^1_{S^3}(\chi,\theta,\varphi)$ in (\ref{CNFRM_LPL}), i.e. its discrete eigenvalues, $(K+1)^2$, acquire meaning of scale dimensions (canonical plus anomalous), and the  Green function  for $\chi=\pi$ becomes proportional to  $\Gamma_{\mbox{cusp}}(\chi) $.
Same expressions have been calculated in \cite{Belitsky} from the equivalent approach of considering the propagator of a particle on $S^3$, i.e. 
the transition amplitude for such a particle  to go from a point $r$ to a point $r^\prime$ on $S^3$. 

In effect, in showing proportionality between the cusp function in the equation
(\ref{cuspfunction}) and the Green function ${\mathcal G}_\pi(\chi) $ in the above equation (\ref{GSouth_1}) 
allows to interpret  the cusp function as a potential generated by a single color charge located at the South pole of the sphere, i.e.

\begin{equation}
-\frac{q_2}{4\pi^2}\Gamma_{\mbox{cusp}}(\chi)=q_2G_\pi (\chi)=-\frac{q_2}{4\pi^2}\chi\cot\chi, \quad {\bar c}_1=c_1=0.
\end{equation}
A $\Gamma_{\mbox{cusp}}(\chi)$ function calculated at $\chi\to (\chi -\pi)$, becomes $(\chi -\pi)\cot\chi$ \cite{Lovelace}, and thereby proportional to $G_0(\chi)$. In result, the color dipole potential in (\ref{dipole_ctg_1}) can be equivalently re-expressed as,
\begin{eqnarray}
-\frac{q_1}{4\pi^2}\Gamma_{\mbox{cusp}}(\chi -\pi)&-&\frac{q_2}{4\pi^2}\Gamma_{\mbox{cusp}}(\chi)=
q_1 G_0 (\chi) +q_2G_\pi (\chi)\nonumber\\
& =&-\frac{q}{4\pi}\cot \chi,\quad q_1=-q_2=-q.
\label{thatis}
\end{eqnarray} 
Therefore, the conformally symmetric color  dipole potential in (\ref{dipole_ctg_1}) and its associated ${\mathbf E}$ field in (\ref{ENEW}) have been
motivated by cusped Wilson loops. In this manner, the notion of a  ``geometric confinement'' could be introduced as a conformal symmetry motivated color neutrality on a closed $S^3$  space, an option in which the color neutrality in QCD appearing  as a consequence of  the color gauge $SU(3)_c/Z_3$ dynamics may express itself in the external space.

Ours is a constructive definition as it  predicted the potential generated by a color-anti-color pair. Testing this prediction by data
has been  the goal of the section 4. There, we showed that the experimental data  on  the  meson spectra with excitations  above
$\sim$ 1400 MeV and below $\sim$2350 MeV,  closely followed  the patterns of the spectra in (\ref{mass_bound}), (\ref{Mas_idtfct_res}), generalized by (\ref{RM_V1}), motivated by (\ref{thatis}).

\begin{table} 
 \begin{center}
     \resizebox{0.85\textwidth}{!}
{\begin{minipage}{\textwidth}
\begin{tabular}{|c|c|c|c|c|}  
\hline 
trajectory &A(R)& -B(R)  & C & $\sigma_M^2$
\\ \hline \hline 
$\pi $ & 0.10964992 GeV$^2$ & 
-1.57653471 GeV$^2$ & 1.48636457 GeV$^2$ &0.0714330 GeV$^2$  \\ 
\hline 

$f_0$ & 0.10964992 GeV$^2$ & -1.0434231 GeV$^2$ &
1.18377318 GeV$^2$ &0.1035338 GeV$^2$ \\ 
\hline 
$\eta $ &0.10964992 GeV$^2$ & -1.2949872 GeV$^2$  & 1.48548786 GeV$^2$ &0.1010973 GeV$^2$ \\
\hline 
$a_0$ &0.10964992 GeV$^2$ & -3.7453673 GeV$^2$ & -2.41226501 GeV$^2$ &0.0666480 GeV$^2$ \\
\hline
\end{tabular} 
\caption{Parameters of the least square fit to the meson resonance masses on the trajectories indicated to the very left by the mass formula in (\ref{energies_tRM}). The last column contains the standard deviation, denoted by $\sigma^2_M$.  
} 
\end{minipage}}
  \end{center}
\label{Table1} 
\end{table}

\begin{table} 
 \begin{center}
  \resizebox{0.85\textwidth}{!}
{\begin{minipage}{\textwidth}
\begin{tabular}{|c|c|c|c|c|c|c|c|c|c|c|c|c|}  
\hline 
meson     & $\pi$   & $b_1$  & $b_1$ &$b_3$  & $\pi_4$ & $b_5$ &$\eta_2$ & $h_3$ &$\eta_4$ & $h_5$ & $a_0$  & $a_0$  \\ 
\\ \hline 
prediction & 1516  & 1516  & 1773 & 1773 & 2041  & 2322 & 1526 & 1777 & 2043  &2330  & 1689  & 2275  \\ 
\hline \hline
\end{tabular} 
\caption{Predictions by the fit of  the masses, in MeV, of the 12 missing mesons. 
 Mesons belonging to same pole are supposed to have equal masses.
It has to be remarked  that the formation of some of these mesons may 
be suppressed by some internal dynamics such as proximity to thresholds etc.    
Mesons of equal masses correspond to same $SO(4)$ pole.} 
\end{minipage}}
  \end{center}
\label{Table3} 
\end{table} 

\begin{table} 
 \begin{center}
 \resizebox{0.85\textwidth}{!}
{\begin{minipage}{\textwidth}
\begin{tabular}{|c|c|c||c|}  
\hline 
number of nodes & meson & predicted mass $M^{\mbox{th}}$ & $\left[M^{\mbox{exp}}-M^{\mbox{th}}\right]/  M^{\mbox{exp}}$ \\
\\ \hline \hline 

 0  & $\pi (139.57)$  & 139.57& 0.00\%    \\ 
\hline 

 0 &$b_1(1235)$   & 1237.26738 &0.18\% \\ 
\hline 
0& $\pi_2(1670)$ & 1515.92985 & -9.23\% \\
\hline 
1 & $\pi (1300)$ & 1237.26738  & -4.83 \%\\
\hline
1& $\pi_2(1880)$ & 1772.63359& -5.71\%   \\
\hline
 1 & $b_3(2030)$ & 2040.72319  & 0.53\% \\
\hline
1 & $\pi_4 (2250)$ & 2321.63067 & 3.18\% \\
\hline
2 &$\pi_2(2100)$ & 2040.72319 & -2.82\% \\
\hline
2 & $b_3(2245)$ & 2321.63067 &  3.41\%\\
\hline
3 & $\pi (1800)$ &1772.63359 & -1.52\%\\
\hline
3 & $b_1(1960)$ & 2040.72319 & 4.12\% \\
\hline
3& $\pi_2(2285)$ & 2321.63067 & 1.60\% \\
\hline
4& $\pi (2070)$ & 2040.72319 & -1.41\%\\
\hline
4&$b_1(2240)$ & 2321.63067 & 3.64\% \\
\hline
5 & $\pi (2360)$ & 2321.63067 & -1.63\%\\
\end{tabular} 
\caption{Illustrative example for the precision of the data fit by the theoretically predicted  $SO(4)$ poles for the particular case of the $\pi$ trajectory.
The measured masses,  $M^{\mbox{exp}}$,  (the number inside the parenthesis of the meson notation) are compared to the predictions of this work,
$M^{\mbox{th}}$.  The relative deviation,$\left[M^{\mbox{exp}}-M^{\mbox{th}}\right]/M^{\mbox{exp}}$ , in percentages, is also given. The fist column contains the number of nodes of the wave functions defining a sub trajectory of the conformal trajectory.
} 
\end{minipage}}
\end{center}
\end{table}

\begin{table} 
 \begin{center}
     \resizebox{0.85\textwidth}{!}
{\begin{minipage}{\textwidth}
\begin{tabular}{|c|c|c|c|}  
\hline 
$I^G\left(\ell ^{PC}\right)$&meson & ref.~\cite{Hess} &  this work
\\ \hline \hline 
$0^+(0^{-+})$&$\eta (1440)$ & 1 379 MeV & 
  1372 MeV \\ 
\hline 

$0^+(0^{-+})$&$\eta (1295)$ & 1428 MeV  & 1115 MeV \\ 
\hline 
$0^+(0^{-+})$&$\eta ( 1760)$ & 1671 MeV & 1674 MeV  \\
\hline 
$0^-(1^{--})$&$\omega (1420)$ & 1389 MeV & 1398  MeV \\
\hline
$0^-(1^{--})$&$\omega (1650)$ & 1639 MeV & 1679 MeV  \\
\hline
\end{tabular} 
\caption{ Comparison between meson masses predicted by Ref.~\cite{Hess} and  the present work. The mesons of the lowest masses have been omitted from
the comparison because we fix their masses  to the empirical values and do not consider them as predictions.    
} 
\end{minipage}}
  \end{center}
\label{Table4} 
\end{table}

\newpage

\begin{flushleft}
{\Large \bf Addendum to ``Modelling duality between bound and resonance meson spectra by means of free quantum motions on the de Sitter space time $dS_4$'',
Eur. Phys.J. A (2016) {\bf 52}:210}
\end{flushleft}

\vspace{0.35cm}

\begin{center}
 M. Kirchbach$^1$,
C.\ B.\ Compean$^2$
\end{center}

\begin{center}
$^1$Instituto de F{\'{i}}sica, UASLP,\\
Av. Manuel Nava 6, Zona Universitaria,\\
San Luis Potos{\'{i}}, S.L.P. 78290, M\'exico\\
$^2$
 Instituto Tecnol\'ogico de San Luis Potos\'{\i},\\
  Av. Tecnol\'ogico S/N col. UPA, Soledad de Graciano S\'anchez, S.L.P. 78437, M\'exico
\end{center}

\begin{flushleft}
{\bf Abstract:}In the article under discussion the analysis of the spectra of the
unflavored mesons lead us to some intriguing insights into the possible geometry of
space-time outside the causal Minkowski  light cone and into the nature of 
strong interactions. In applying the potential theory concept  of geometrization of interactions, we showed that the meson masses are best described by a confining potential composed by the centrifugal barrier on the three dimensional spherical space, $S^3$,  and of a charge-dipole potential constructed from the Green function to the $S^3$ Laplacian. The dipole potential emerged  in view of the fact that $S^3$ does not support single-charges without violation of the Gauss theorem and the superposition principle, thus providing a natural stage  for the description of the general phenomenon of confined charge-neutral systems.  However, in the original article we did not relate the charge-dipoles on $S^3$ to the color neutral  mesons, and did not express the magnitude of the confining dipole potential in terms of the strong coupling $\alpha_S$ and the number of colors, $N_c$, the subject of the addendum. To the amount $S^3$ can be thought of as the unique closed space-like  geodesic of a four-dimensional de Sitter space-time, $dS_4$, we hypothesized the space-like region outside the causal Einsteinian light cone (it describes virtual processes, among them interactions) as the  $(1+4)$ dimensional subspace of the conformal $(2+4)$ space-time, foliated with $dS_4$ hyperboloids, and in this way assumed relevance of  $dS_4$ special relativity for strong interaction processes. The potential designed in this way predicted meson spectra of conformal degeneracy patterns, and in accord with the experimental observations. We now extract the $\alpha_s$ values in the infrared from data on meson masses. The results obtained are compatible with the  $\alpha_s$ estimates provided by other approaches. 
\end{flushleft}
\vspace{0.35cm}
\centerline{PACS: { {12.39.Jh} {(Non relativistic quark models)}, {14.40.Be} {(Light mesons)}, {03.65.Fd} {(Algebraic methods)}, {02.30.Ik} {(Integrable systems)}}}
\vspace{0.35cm}

{}For the purpose of explaining the  spectra of the unflavored mesons with masses below $\sim 2300$ MeV, we modelled 
in [1] the QCD dynamics by means of differential equations of second order of the Sturm-Liouville type with potentials.  Towards this goal we were
guided by  the concepts of potential theory [2] according to which instantaneous interactions
are composed by the centrifugal barriers on surfaces complemented by terms calculated from the Green functions of the corresponding Laplacians. For example, the famous Coulomb potential is obtained from combining the centrifugal barrier, $\ell(\ell+1)/r^2$ of the flat three dimensional space, with the (properly parametrized) solution, $\sim 1/r$, of the Laplace equation, $\nabla^2f(r)=0$. 
The main result of  [1] is that the spectra of the unflavored mesons with masses below $\sim 2300$ MeV require for their adequate quantum mechanical description   the centrifugal potentials to be those on the open time-like hyperbolic, or closed space-like hyper-spherical  $dS_4$ geodesic, in depending whether the mesons have been treated as resonances transmitted through the P\"oschl-Teller barrier, or, as states bound within the trigonometric Scarf potential. 
In [1] we focused on the case of the unique closed space-like geodesic, the three dimensional hyper spherical space $S^3$, located at the equator of the four-dimensional hyperboloid of one-sheet representing $dS_4$. We drew attention to the textbook fact known from  
[3] that no consistent single-charge definition can be formulated on $S^3$ , and that due to the innate charge neutrality of the hypersphere, only charge-neutral systems, such as  dipole sources, can  generate potentials consistent with the Gauss theorem and the superposition principle on $S^3$.  
In employing the  concepts of potential theory [2], we derived in the equation (A.20) in [1]
 such a potential from the Green function of the conformal Laplacian on $S^3$, obtaining it as $-q/(4\pi )  \cot \chi$, with $q$ standing for a generic charge, and $\chi \in [0,\pi]$ denoting  the second polar angle parameterizing $S^3$.
In the present addendum we explore consequence of identifying the generic charge with the color charge, $g$, in QCD.
This is justified for the following two reasons. Firstly, the $S^3$ hypersphere  provides a stage suited  for the description of the general phenomenon of charge-neutral systems confined to a conformally symmetric space, a phenomenon so far known only for the color charges in QCD, and secondly,  our Sturmian modelling of the  QCD dynamics requires the potentials to be  constructed from  that very same  $S^3$ kinematic, and the Green function to the corresponding Laplacian.  Finally we  notice that instantaneous interactions, in not allowing for time orderings, represent virtual processes ruled by the space-like region outside the causal Minkowski light cone, an observation suggestive of considering $dS_4$ as the possible geometry of the space-like region relevant for strong interactions. In order to conserve conventional special relativity in the time-like region and the dimensionality of the Minkowski light cone, the latter has to be attached to a local observer on $dS_4$ and described as  a $dS_4$ causal patch, i.e. as a $dS_4$ intersection by planes parallel to the time axis, a view that takes one to the $dS_4$ special relativity hypothesis. The $dS_4$ hyperboloids can be viewed as foliation of the $(1+4)$ subspace of a conformal space-time, $(2+4)$. As a comparison, the usual Minkowski space-like region is foliated by $dS_3$ hyperboloids, quantum motions on which give raise to potentials completely unsuited as strong interactions. In this sense, hadrons description seems to favor de Sitter,  (see 
[4] for a historical overview and an extensive  bibliography on de Sitter special relativity) over Einsteinian special relativity.  Because of the conformal symmetry of $dS_4$, the charge confining potential, $-q/(4\pi)\cot\chi$,  discussed above, is conformal too. 
Upon  identifying the generic charge $q$ with the color charge, $g$, by setting $q=g$,  the potential energy of a charge $g$ within this potential becomes, $-g^2/(4\pi) \cot\chi$. In now taking into
account the multiplicity of the color charge, $N_c$, i.e. the number of possible charges in the source, and denoting
$g^2/(4\pi)$ by the QCD strong coupling $\alpha_s$, the following color-dipole 
interaction is obtained,
\begin {equation}
 V(\chi)=-\alpha_s N_c\cot\chi, \quad \chi\in[0,\pi].
\label{ourPt}
\end{equation}
The latter expression can be independently verified from calculating Wilson
cusped loops  on $S^3$. Indeed, it follows from the eqs. (B.2) and (B.7) in [1]
that a potential generated by a single charge source  located at the South
pole in accord with (B.22)  ($N_c$ times replicated) is
perceived by a local color charge  as
\begin{equation}
V_{South}(\chi)=-\frac{\alpha_s}{\pi}N_c\chi \cot\chi.
\label{Cusp_South}
\end{equation}
For a source located at the North pole, one has instead,

\begin{equation}
V_{North}(\chi) =
\frac{\alpha_s}{\pi}N_c(\pi -\chi)\cot\chi.
\label{Cusp_North}
\end{equation}
The innate charge neutrality on the hypersphere requires, as explained at 
lengths in the article,  a color
 interaction in the dipole form according to,
\begin{equation}
V_{South}(\chi)-V_{North}(\chi)=V(\chi)=-\alpha_sN_c\cot\chi,
\label{Herz10}
\end{equation}
an expression identical to our equation  (\ref{ourPt}) from above.
In this fashion, mesons can be visualized (if needed) as quantum light quarkish color-anti-color ``dumbbells'' in free  4D rotations around their  mass centers,
with the ends ``tracing''  great circles of a 3D hypersphere, a motion  perturbed  by the color dipole potential in (\ref{Herz10}), generated by another 4D color--anti-color quantum  ``dumbbell'', a glueball, sufficiently heavy to remain static to leading order. 
In this way, the data fit performed by us, and  including  the parameter $b$ 
in the equations  (51)-(52) relates to  $\alpha_s N_c$  as,
\begin{equation}
2b=\alpha_sN_c, \quad b=\sqrt{B(R)/A(R)},
\label{alpha_fit}
\end{equation}
with $B(R)$ from Table 1 in [1]. Notice however that due to  a regrettable typo in the printed journal version, the negative
signs of $B(R)$ in the third column of this Table there have all to be read as positive (here corrected). All the other entries in the table remain same.
As a reminder, the net conformal  color-dipole confining potential, here denoted by ${\mathcal V}_{CCD}(\chi)$,  derived in the equation (A.17) in [1]reads,
\begin{eqnarray}
V_{\mbox{CCD}}(\chi)&=&\frac{\ell (\ell +1)}{\sin^2\chi} -2b\cot \chi, \quad \chi\in [0,\pi],
\end{eqnarray}
with $2b$ from (\ref{alpha_fit}). In the present parametrization, the mass formula in the equations (51)-(52) in [1] becomes,
\begin{eqnarray}
M^2&=&A(R)(K+1)^2 -\frac{B(R)}{(K+1)^2} +C,\nonumber\\
A(R)=\frac{\hbar^2c^2}{R^2}, && B(R)=\frac{\alpha^2_sN^2_c\hbar^2 c^2}{4R^2}, \,\, R=0.58\, \mbox{fm}.
\label{mss_fla}
\end{eqnarray} 
The equation (\ref{mss_fla}) allows one to extract the value of the strong
coupling from the meson spectra. In so doing we find for the pion excitations
$\alpha_s/\pi =0.8$, for the $f_0$ excitations we obtain $\alpha_s/\pi =0.65$, 
while for the $\eta$, and $a_0$  meson spectra we find $\alpha_s/\pi =0.73$, and $\alpha_s/\pi=1.08$, respectively.
However,  the $a_0$ spectrum has been fitted in [1] without inclusion of the  ``missing'' state around $500$ MeV, whose formation might have been suppressed by some dynamical reasons,  a circumstance that  amounted to a $B(R)$ value much larges in comparison  to the rest. Had we instead included this state into the fit, and assumed $500$ MeV for its mass, the $\alpha_s/\pi $ value would have come out same as for the $f_0$ meson. We consider this value less reliable than the other three. Notice that
the lightest particle, the pion, predicts  the notably largest $\alpha_s$ value, as it
should be,  given that the strong coupling constant grows  at low energies.
The $\alpha_s/\pi $ values extracted from the meson spectra, in remaining 
notably below 1, are in accord  with data extracted from the spin structure 
function and presented in ref. [15] in [1],  and of the order calculated  in [5],[6].
Finally we like to point out that from the twelve ``missing''  excited mesons, predicted in [1],three are located on the leading $\pi (139)$ trajectory, another one is from the $\pi (1300)$ trajectory, and two more
are absent from the second accompanying trajectory. As regarding the four missing states from the $\eta$ family, they all are  located on the leading trajectory. In this way, ten out of twelve  missing mesons are needed for completing the excitations in the anomaly affected  sector of the pseudo-scalar mesons.
As a comparison, in the anomaly-free scalar meson sector, only two $a_0$ meson resonances  are missing, not counting the absent ground state.
It is important to know whether the absence of the above mentioned ten pseudo-scalar mesons is circumstantial, or  due to
interference effects between the anomaly and the conformal symmetry, a subject requiring more profound studies.

\vspace{0.35cm}

\noindent
[1].  M. Kirchbach and C. B. Compean, Eur.\ Phys.\ J.\ A (2016) {\bf 52}:210.\\

\noindent
[2]. O.\ D.\ Kellogg, {\it Foundations of Potential Theory} (Dover, New York, 1953).\\

\noindent
[3]. L.\ D.\  Landau and  E.\ M.\ Lifschitz, {\it The Classical  Theory of Fields}, Vol. 2 of A Course of Theoretical Physics,
 3d edition (Pergamon Press 1971) p.335. \\

\noindent
[4].  https://en.wikipedia.org/wiki/De\_Sitter\_invariant\_special\_relativity\\

\noindent
[5].  T.\ Gehrman, M.\ Jaquier, and G.\ Luisoni, Eur.\ Phys.\ J.\ C {\bf 67}, 57 (2010).\\

\noindent
[6].  A.\ C.\ Aguilar, D.\ Binosi, J.\ Papavassiliou, and J.\ Rodriguez-Quintero,
Phys.\ Rev.\ D {\bf 80}, 085018 (2009).

\end{document}